\documentclass[12pt]{article}

\usepackage{amsmath}
\usepackage{stmaryrd}
\usepackage{subfigure}
\usepackage{empheq}
\usepackage{xcolor}
\usepackage{appendix,epic,eepic,amsmath,amsthm,amssymb,epsfig,cite,indentfirst,oldgerm}
\renewcommand{\theequation}{\arabic{equation}}
\newcommand{\EQ}{\begin{equation}}
\newcommand{\EN}{\end{equation}}

\newcommand{\ket}[1]{\left|#1\right\rangle}      

\newcommand{\bear}{\begin{eqnarray}}
\newcommand{\ear}{\end{eqnarray}}
\newcommand{\bt} { \begin{tabular} }
\newcommand{\et}{ \end{tabular} }
\newcommand{\bc} { \begin{center} }
\newcommand{\ec}{ \end{center} }

\newcommand{\btb} { \begin{table} }
\newcommand{\etb}{ \end{table} }

\begin{document}

\topmargin 0pt
\oddsidemargin 5mm
\newcommand{\NP}[1]{Nucl.\ Phys.\ {\bf #1}}
\newcommand{\PL}[1]{Phys.\ Lett.\ {\bf #1}}
\newcommand{\NC}[1]{Nuovo Cimento {\bf #1}}
\newcommand{\CMP}[1]{Comm.\ Math.\ Phys.\ {\bf #1}}
\newcommand{\PR}[1]{Phys.\ Rev.\ {\bf #1}}
\newcommand{\PRL}[1]{Phys.\ Rev.\ Lett.\ {\bf #1}}
\newcommand{\MPL}[1]{Mod.\ Phys.\ Lett.\ {\bf #1}}
\newcommand{\JETP}[1]{Sov.\ Phys.\ JETP {\bf #1}}
\newcommand{\TMP}[1]{Teor.\ Mat.\ Fiz.\ {\bf #1}}

\renewcommand{\thefootnote}{\fnsymbol{footnote}}

\newpage
\setcounter{page}{0}
\begin{titlepage}
\begin{flushright}

\end{flushright}
\begin{center}
{\large  On the equivalence between $n$-state spin and vertex models on the square lattice.} \\
\vspace{0.5cm}
{\large M.J. Martins } \\
\vspace{0.15cm}
{\em Universidade Federal de S\~ao Carlos\\
Departamento de F\'{\i}sica \\
C.P. 676, 13565-905, S\~ao Carlos (SP), Brazil\\}
\vspace{0.35cm}
\end{center}
\begin{abstract}
In this paper we investigate a correspondence 
among
spin and vertex models with the same number 
of local states on the square lattice with 
toroidal boundary conditions.
We argue that the partition functions of 
an arbitrary $n$-state 
spin model
and of a certain specific $n$-state vertex 
model coincide 
for finite lattice sizes. The equivalent vertex model 
has $n^3$ non-null Boltzmann weights and their 
relationship with the 
edge weights of the spin model is explicitly 
presented. In particular,
the Ising model in a magnetic field is mapped 
to an eight-vertex model whose
weights configurations combine both even and odd 
number of incoming 
and outcoming arrows at a vertex. We have studied the
Yang-Baxter algebra for such mixed eight-vertex model when
the weights are invariant under arrows reversing. We find that while the
Lax operator lie on the same elliptic curve of the even eight-vertex
model the respective $\mathrm{R}$-matrix can not be presented in terms
of the difference of two rapidities. 
We also argue that the spin-vertex  equivalence may be used to imbed an integrable
spin model in the realm of the quantum inverse scattering framework. As
an example, we show how to determine the $\mathrm{R}$-matrix of the 27-vertex model 
equivalent to a
three-state spin model devised by Fateev and Zamolodchikov.
\end{abstract}
\centerline{Keywords: Spin and Vertex models, Yang-Baxter equations, Spin chain }
\centerline{May~~2024}
\end{titlepage}


\pagestyle{empty}

\newpage

\pagestyle{plain}
\pagenumbering{arabic}

\renewcommand{\thefootnote}{\arabic{footnote}}
\newtheorem{proposition}{Proposition}
\newtheorem{pr}{Proposition}
\newtheorem{remark}{Remark}
\newtheorem{re}{Remark}
\newtheorem{theorem}{Theorem}
\newtheorem{theo}{Theorem}

\def\ll{\left\lgroup}
\def\rr{\right\rgroup}

\newtheorem{Theorem}{Theorem}[section]
\newtheorem{Corollary}[Theorem]{Corollary}
\newtheorem{Proposition}[Theorem]{Proposition}
\newtheorem{Conjecture}[Theorem]{Conjecture}
\newtheorem{Lemma}[Theorem]{Lemma}
\newtheorem{Example}[Theorem]{Example}
\newtheorem{Note}[Theorem]{Note}
\newtheorem{Definition}[Theorem]{Definition}

\section{Introduction}

In statistical mechanics systems are sometimes modeled by 
lattice models in which the set of possible 
microstates are specified by placing local spin variables
along the interacting sites of the lattice. 
An energy interaction or
equivalently a Boltzmann weight is
then assigned to each possible microstate configuration according
to the specific lattice model. In this paper we focus on a square
lattice of size $L \times L$ with periodic boundary 
conditions in the horizontal and vertical directions. Here we shall 
also consider
that the microstates are described by discrete spin variables 
assuming $n$ possible values.  
One of the simplest type of such model is when the 
spin variable $\sigma_{i,j}$ sits on the lattice site $(i,j)$ 
and the energy interactions involve only nearest neighbors 
sites configurations. These systems are called spin models and the
simplest prototype is a two state system known as the 
Ising model \cite{MOM,MCT} whose
partition function is,
\begin{equation}
\label{PARISI}
Z_{\mathrm{Ising}}(L)= \sum_{<\sigma_{i,j}>} \mathrm{exp}\Big [\beta \sum_{i,j=1}^{L} \left( J_h \sigma_{i,j} \sigma_{i,j+1} +J_v 
\sigma_{i,j} \sigma_{i+1,j} +H \sigma_{i,j} \right ) \Big ]
\end{equation}
where the sum $<\sigma_{i,j}>$ is over all the two-state spins variables 
$\sigma_{i,j}=\pm$ of the  lattice and
$\beta=\frac{1}{k_B T}$ is the thermal factor.
The couplings $J_h$ and $J_v$ 
correspond 
to the interaction 
energies in the horizontal and vertical
lattice directions and $H$ represents an external magnetic field. This model in the absence of a magnetic field 
was solved by Onsager who has computed exactly
the respective free energy \cite{ONSA1}.

Another important family of lattice models are vertex models in which the
spin variables sit on the four links of given site of the square lattice. 
They have emerged 
in the context of the residual entropy of the ice and in certain phase 
transition exhibited by
hydrogen-bonded crystals \cite{PAU,SLA}. 
The statistical 
configurations are characterized by the two possible positions of the
hydrogens which usually are indicated by incoming and outcoming arrows placed
along the links \cite{LIWU}. One may also use an alternative
description of the statistical configurations since there exists a direct 
correspondence between arrow configurations and a two state
spin variable denoted here by the states $\pm$.
This gives rise to a two state vertex model and 
if no restriction is imposed to the hydrogen atoms positions we have
on the square lattice sixteen possible vertex weights configurations. These 
vertex configurations are shown in Fig.(\ref{Fig16V}) using both the arrow
configurations and the spin $\pm$ variables.
\setlength{\unitlength}{2500sp}
\begin{figure}[ht]
\begin{center}
\begin{picture}(8000,2000)
{\put(-2200,900){\line(1,0){1400}}}
{\put(-600,900){\line(1,0){1400}}}
{\put(1000,900){\line(1,0){1400}}}
{\put(2600,900){\line(1,0){1400}}}
{\put(4200,900){\line(1,0){1400}}}
{\put(5800,900){\line(1,0){1400}}}
{\put(7400,900){\line(1,0){1400}}}
{\put(9000,900){\line(1,0){1400}}}
{\put(-2200,-1200){\line(1,0){1400}}}
{\put(-600,-1200){\line(1,0){1400}}}
{\put(1000,-1200){\line(1,0){1400}}}
{\put(2600,-1200){\line(1,0){1400}}}
{\put(4200,-1200){\line(1,0){1400}}}
{\put(5800,-1200){\line(1,0){1400}}}
{\put(7400,-1200){\line(1,0){1400}}}
{\put(9000,-1200){\line(1,0){1400}}}
{\put(-1450,-500){\line(0,-1){1400}}}
{\put(138,-500){\line(0,-1){1400}}}
{\put(1740,-500){\line(0,-1){1400}}}
{\put(3338,-500){\line(0,-1){1400}}}
{\put(4940,-500){\line(0,-1){1400}}}
{\put(6540,-500){\line(0,-1){1400}}}
{\put(8140,-500){\line(0,-1){1400}}}
{\put(9740,-500){\line(0,-1){1400}}}
{\put(-1450,1600){\line(0,-1){1400}}}
{\put(138,1600){\line(0,-1){1400}}}
{\put(1740,1600){\line(0,-1){1400}}}
{\put(3338,1600){\line(0,-1){1400}}}
{\put(4940,1600){\line(0,-1){1400}}}
{\put(6540,1600){\line(0,-1){1400}}}
{\put(8140,1600){\line(0,-1){1400}}}
{\put(9740,1600){\line(0,-1){1400}}}
{\put(-1800,890){\makebox(0,0){\fontsize{12}{14}\selectfont $\color{red} >$}}}
{\put(-215,890){\makebox(0,0){\fontsize{12}{14}\selectfont $\color{red} <$}}}
{\put(1385,890){\makebox(0,0){\fontsize{12}{14}\selectfont $\color{red} >$}}}
{\put(2985,890){\makebox(0,0){\fontsize{12}{14}\selectfont $\color {red} <$}}}
{\put(4585,890){\makebox(0,0){\fontsize{12}{14}\selectfont $\color{red} >$}}}
{\put(6185,890){\makebox(0,0){\fontsize{12}{14}\selectfont $\color{red} <$}}}
{\put(7785,890){\makebox(0,0){\fontsize{12}{14}\selectfont $\color{red} <$}}}
{\put(9350,890){\makebox(0,0){\fontsize{12}{14}\selectfont $\color{red} >$}}}
{\put(-2140,1010){\makebox(0,0){\fontsize{10}{12}\selectfont $\color{blue} +$}}}
{\put(-530,1010){\makebox(0,0){\fontsize{10}{12}\selectfont $\color{blue} -$}}}
{\put(1070,1010){\makebox(0,0){\fontsize{10}{12}\selectfont $\color{blue} +$}}}
{\put(2650,1010){\makebox(0,0){\fontsize{10}{12}\selectfont $\color{blue} -$}}}
{\put(4250,1010){\makebox(0,0){\fontsize{10}{12}\selectfont $\color{blue} +$}}}
{\put(5850,1010){\makebox(0,0){\fontsize{10}{12}\selectfont $\color{blue} -$}}}
{\put(7450,1010){\makebox(0,0){\fontsize{10}{12}\selectfont $\color{blue} -$}}}
{\put(9015,1010){\makebox(0,0){\fontsize{10}{12}\selectfont $\color{blue} +$}}}
{\put(-1800,-1200){\makebox(0,0){\fontsize{12}{14}\selectfont $\color{red} >$}}}
{\put(-215,-1200){\makebox(0,0){\fontsize{12}{14}\selectfont $\color{red} <$}}}
{\put(1385,-1200){\makebox(0,0){\fontsize{12}{14}\selectfont $\color{red} >$}}}
{\put(2985,-1200){\makebox(0,0){\fontsize{12}{14}\selectfont $\color{red} <$}}}
{\put(4585,-1200){\makebox(0,0){\fontsize{12}{14}\selectfont $\color{red} >$}}}
{\put(6185,-1200){\makebox(0,0){\fontsize{12}{14}\selectfont $\color{red} <$}}}
{\put(7785,-1200){\makebox(0,0){\fontsize{12}{14}\selectfont $\color{red} <$}}}
{\put(9350,-1200){\makebox(0,0){\fontsize{12}{14}\selectfont $\color{red} >$}}}
{\put(-2140,-1100){\makebox(0,0){\fontsize{10}{12}\selectfont $\color{blue} +$}}}
{\put(-530,-1100){\makebox(0,0){\fontsize{10}{12}\selectfont $\color{blue} -$}}}
{\put(1070,-1100){\makebox(0,0){\fontsize{10}{12}\selectfont $\color{blue} +$}}}
{\put(2650,-1100){\makebox(0,0){\fontsize{10}{12}\selectfont $\color{blue} -$}}}
{\put(4250,-1100){\makebox(0,0){\fontsize{10}{12}\selectfont $\color{blue} +$}}}
{\put(5850,-1100){\makebox(0,0){\fontsize{10}{12}\selectfont $\color{blue} -$}}}
{\put(7450,-1100){\makebox(0,0){\fontsize{10}{12}\selectfont $\color{blue} -$}}}
{\put(9015,-1100){\makebox(0,0){\fontsize{10}{12}\selectfont $\color{blue} +$}}}
{\put(-1140,890){\makebox(0,0){\fontsize{12}{14}\selectfont $\color{red} >$}}}
{\put(460,890){\makebox(0,0){\fontsize{12}{14}\selectfont $\color{red} <$}}}
{\put(2060,890){\makebox(0,0){\fontsize{12}{14}\selectfont $\color{red} >$}}}
{\put(3660,890){\makebox(0,0){\fontsize{12}{14}\selectfont $\color{red} <$}}}
{\put(5260,890){\makebox(0,0){\fontsize{12}{14}\selectfont $\color{red} <$}}}
{\put(6860,890){\makebox(0,0){\fontsize{12}{14}\selectfont $\color{red} >$}}}
{\put(8460,890){\makebox(0,0){\fontsize{12}{14}\selectfont $\color{red} >$}}}
{\put(10060,890){\makebox(0,0){\fontsize{12}{14}\selectfont $\color{red} <$}}}
{\put(-920,1010){\makebox(0,0){\fontsize{10}{12}\selectfont $\color{blue} +$}}}
{\put(700,1010){\makebox(0,0){\fontsize{10}{12}\selectfont $\color{blue} -$}}}
{\put(2300,1010){\makebox(0,0){\fontsize{10}{12}\selectfont $\color{blue} +$}}}
{\put(3900,1010){\makebox(0,0){\fontsize{10}{12}\selectfont $\color{blue} -$}}}
{\put(5500,1010){\makebox(0,0){\fontsize{10}{12}\selectfont $\color{blue} -$}}}
{\put(7100,1010){\makebox(0,0){\fontsize{10}{12}\selectfont $\color{blue} +$}}}
{\put(8700,1010){\makebox(0,0){\fontsize{10}{12}\selectfont $\color{blue} +$}}}
{\put(10300,1010){\makebox(0,0){\fontsize{10}{12}\selectfont $\color{blue} -$}}}
{\put(-1140,-1200){\makebox(0,0){\fontsize{12}{14}\selectfont $\color{red} >$}}}
{\put(460,-1200){\makebox(0,0){\fontsize{12}{14}\selectfont $\color{red} <$}}}
{\put(2060,-1200){\makebox(0,0){\fontsize{12}{14}\selectfont $\color{red} >$}}}
{\put(3660,-1200){\makebox(0,0){\fontsize{12}{14}\selectfont $\color{red} <$}}}
{\put(5260,-1200){\makebox(0,0){\fontsize{12}{14}\selectfont $\color{red} <$}}}
{\put(6860,-1200){\makebox(0,0){\fontsize{12}{14}\selectfont $\color{red} >$}}}
{\put(8460,-1200){\makebox(0,0){\fontsize{12}{14}\selectfont $\color{red} >$}}}
{\put(10060,-1200){\makebox(0,0){\fontsize{12}{14}\selectfont $\color{red} <$}}}
{\put(-920,-1100){\makebox(0,0){\fontsize{10}{12}\selectfont $\color{blue} +$}}}
{\put(700,-1100){\makebox(0,0){\fontsize{10}{12}\selectfont $\color{blue} -$}}}
{\put(2300,-1100){\makebox(0,0){\fontsize{10}{12}\selectfont $\color{blue} +$}}}
{\put(3900,-1100){\makebox(0,0){\fontsize{10}{12}\selectfont $\color{blue} -$}}}
{\put(5500,-1100){\makebox(0,0){\fontsize{10}{12}\selectfont $\color{blue} -$}}}
{\put(7100,-1100){\makebox(0,0){\fontsize{10}{12}\selectfont $\color{blue} +$}}}
{\put(8700,-1100){\makebox(0,0){\fontsize{10}{12}\selectfont $\color{blue} +$}}}
{\put(10300,-1100){\makebox(0,0){\fontsize{10}{12}\selectfont $\color{blue} -$}}}
{\put(-1450,1240){\makebox(0,0){\fontsize{12}{14}\selectfont $\color{red} \wedge$}}}
{\put(140,1240){\makebox(0,0){\fontsize{12}{14}\selectfont $\color{red} \vee$}}}
{\put(1740,1240){\makebox(0,0){\fontsize{12}{14}\selectfont $\color{red} \vee$}}}
{\put(3340,1240){\makebox(0,0){\fontsize{12}{14}\selectfont $\color{red} \wedge$}}}
{\put(4940,1240){\makebox(0,0){\fontsize{12}{14}\selectfont $\color{red} \wedge$}}}
{\put(6540,1240){\makebox(0,0){\fontsize{12}{14}\selectfont $\color{red} \vee$}}}
{\put(8150,1240){\makebox(0,0){\fontsize{12}{14}\selectfont $\color{red} \wedge$}}}
{\put(9750,1240){\makebox(0,0){\fontsize{12}{14}\selectfont $\color{red} \vee$}}}
{\put(-1570,1520){\makebox(0,0){\fontsize{10}{12}\selectfont $\color{blue} +$}}}
{\put(20,1520){\makebox(0,0){\fontsize{10}{12}\selectfont $\color{blue} -$}}}
{\put(1620,1520){\makebox(0,0){\fontsize{10}{12}\selectfont $\color{blue} -$}}}
{\put(3220,1520){\makebox(0,0){\fontsize{10}{12}\selectfont $\color{blue} +$}}}
{\put(4820,1520){\makebox(0,0){\fontsize{10}{12}\selectfont $\color{blue} +$}}}
{\put(6420,1520){\makebox(0,0){\fontsize{10}{12}\selectfont $\color{blue} -$}}}
{\put(8030,1520){\makebox(0,0){\fontsize{10}{12}\selectfont $\color{blue} +$}}}
{\put(9630,1520){\makebox(0,0){\fontsize{10}{12}\selectfont $\color{blue} -$}}}
{\put(-1450,-860){\makebox(0,0){\fontsize{12}{14}\selectfont $\color{red} \vee$}}}
{\put(140,-860){\makebox(0,0){\fontsize{12}{14}\selectfont $\color{red} \wedge$}}}
{\put(1740,-860){\makebox(0,0){\fontsize{12}{14}\selectfont $\color{red} \wedge$}}}
{\put(3340,-860){\makebox(0,0){\fontsize{12}{14}\selectfont $\color{red} \vee$}}}
{\put(4940,-860){\makebox(0,0){\fontsize{12}{14}\selectfont $\color{red} \vee$}}}
{\put(6540,-860){\makebox(0,0){\fontsize{12}{14}\selectfont $\color{red} \wedge$}}}
{\put(8150,-860){\makebox(0,0){\fontsize{12}{14}\selectfont $\color{red} \vee$}}}
{\put(9750,-860){\makebox(0,0){\fontsize{12}{14}\selectfont $\color{red} \wedge$}}}
{\put(-1570,-580){\makebox(0,0){\fontsize{10}{12}\selectfont $\color{blue} -$}}}
{\put(20,-580){\makebox(0,0){\fontsize{10}{12}\selectfont $\color{blue} +$}}}
{\put(1620,-580){\makebox(0,0){\fontsize{10}{12}\selectfont $\color{blue} +$}}}
{\put(3220,-580){\makebox(0,0){\fontsize{10}{12}\selectfont $\color{blue} -$}}}
{\put(4820,-580){\makebox(0,0){\fontsize{10}{12}\selectfont $\color{blue} -$}}}
{\put(6420,-580){\makebox(0,0){\fontsize{10}{12}\selectfont $\color{blue} +$}}}
{\put(8030,-580){\makebox(0,0){\fontsize{10}{12}\selectfont $\color{blue} -$}}}
{\put(9630,-580){\makebox(0,0){\fontsize{10}{12}\selectfont $\color{blue} +$}}}
{\put(-1450,540){\makebox(0,0){\fontsize{12}{14}\selectfont $\color{red} \wedge$}}}
{\put(140,540){\makebox(0,0){\fontsize{12}{14}\selectfont $\color{red} \vee$}}}
{\put(1740,540){\makebox(0,0){\fontsize{12}{14}\selectfont $\color{red} \vee$}}}
{\put(3340,540){\makebox(0,0){\fontsize{12}{14}\selectfont $\color{red} \wedge$}}}
{\put(4940,540){\makebox(0,0){\fontsize{12}{14}\selectfont $\color{red} \vee$}}}
{\put(6540,540){\makebox(0,0){\fontsize{12}{14}\selectfont $\color{red} \wedge$}}}
{\put(8150,540){\makebox(0,0){\fontsize{12}{14}\selectfont $\color{red} \vee$}}}
{\put(9750,540){\makebox(0,0){\fontsize{12}{14}\selectfont $\color{red} \wedge$}}}
{\put(-1330,280){\makebox(0,0){\fontsize{10}{12}\selectfont $\color{blue} +$}}}
{\put(260,280){\makebox(0,0){\fontsize{10}{12}\selectfont $\color{blue} -$}}}
{\put(1860,280){\makebox(0,0){\fontsize{10}{12}\selectfont $\color{blue} -$}}}
{\put(3450,280){\makebox(0,0){\fontsize{10}{12}\selectfont $\color{blue} +$}}}
{\put(5050,280){\makebox(0,0){\fontsize{10}{12}\selectfont $\color{blue} -$}}}
{\put(6650,280){\makebox(0,0){\fontsize{10}{12}\selectfont $\color{blue} +$}}}
{\put(8270,280){\makebox(0,0){\fontsize{10}{12}\selectfont $\color{blue} -$}}}
{\put(9870,280){\makebox(0,0){\fontsize{10}{12}\selectfont $\color{blue} +$}}}
{\put(-1450,-1540){\makebox(0,0){\fontsize{12}{14}\selectfont $\color{red} \wedge$}}}
{\put(140,-1540){\makebox(0,0){\fontsize{12}{14}\selectfont $\color{red} \vee$}}}
{\put(1740,-1540){\makebox(0,0){\fontsize{12}{14}\selectfont $\color{red} \vee$}}}
{\put(3340,-1540){\makebox(0,0){\fontsize{12}{14}\selectfont $\color{red} \wedge$}}}
{\put(4940,-1540){\makebox(0,0){\fontsize{12}{14}\selectfont $\color{red} \vee$}}}
{\put(6540,-1540){\makebox(0,0){\fontsize{12}{14}\selectfont $\color{red} \wedge$}}}
{\put(8150,-1540){\makebox(0,0){\fontsize{12}{14}\selectfont $\color{red} \vee$}}}
{\put(9750,-1540){\makebox(0,0){\fontsize{12}{14}\selectfont $\color{red} \wedge$}}}
{\put(-1330,-1820){\makebox(0,0){\fontsize{10}{12}\selectfont $\color{blue} +$}}}
{\put(260,-1820){\makebox(0,0){\fontsize{10}{12}\selectfont $\color{blue} -$}}}
{\put(1860,-1820){\makebox(0,0){\fontsize{10}{12}\selectfont $\color{blue} -$}}}
{\put(3450,-1820){\makebox(0,0){\fontsize{10}{12}\selectfont $\color{blue} +$}}}
{\put(5050,-1820){\makebox(0,0){\fontsize{10}{12}\selectfont $\color{blue} -$}}}
{\put(6650,-1820){\makebox(0,0){\fontsize{10}{12}\selectfont $\color{blue} +$}}}
{\put(8270,-1820){\makebox(0,0){\fontsize{10}{12}\selectfont $\color{blue} -$}}}
{\put(9870,-1820){\makebox(0,0){\fontsize{10}{12}\selectfont $\color{blue} +$}}}
{\put(-1430,-50){\makebox(0,0){\fontsize{12}{14}\selectfont $w_1$}}}
{\put(160,-50){\makebox(0,0){\fontsize{12}{14}\selectfont $w_2$}}}
{\put(1770,-50){\makebox(0,0){\fontsize{12}{14}\selectfont $w_3$}}}
{\put(3370,-50){\makebox(0,0){\fontsize{12}{14}\selectfont $w_4$}}}
{\put(4970,-50){\makebox(0,0){\fontsize{12}{14}\selectfont $w_5$}}}
{\put(6580,-50){\makebox(0,0){\fontsize{12}{14}\selectfont $w_6$}}}
{\put(8190,-50){\makebox(0,0){\fontsize{12}{14}\selectfont $w_7$}}}
{\put(9780,-50){\makebox(0,0){\fontsize{12}{14}\selectfont $w_8$}}}
{\put(-1430,-2100){\makebox(0,0){\fontsize{12}{14}\selectfont $v_1$}}}
{\put(160,-2100){\makebox(0,0){\fontsize{12}{14}\selectfont $v_2$}}}
{\put(1770,-2100){\makebox(0,0){\fontsize{12}{14}\selectfont $v_3$}}}
{\put(3370,-2100){\makebox(0,0){\fontsize{12}{14}\selectfont $v_4$}}}
{\put(4970,-2100){\makebox(0,0){\fontsize{12}{14}\selectfont $v_5$}}}
{\put(6580,-2100){\makebox(0,0){\fontsize{12}{14}\selectfont $v_6$}}}
{\put(8190,-2100){\makebox(0,0){\fontsize{12}{14}\selectfont $v_7$}}}
{\put(9780,-2100){\makebox(0,0){\fontsize{12}{14}\selectfont $v_8$}}}
\end{picture}
\end{center}
\vspace{2.0cm}
\caption{The sixteen-vertex configurations of the general two state vertex model. In the top row we showed the even eight 
energy weights $w_i$ and in the down row we have
indicated the odd energy weights by $v_i$.}
\label{Fig16V}
\end{figure}
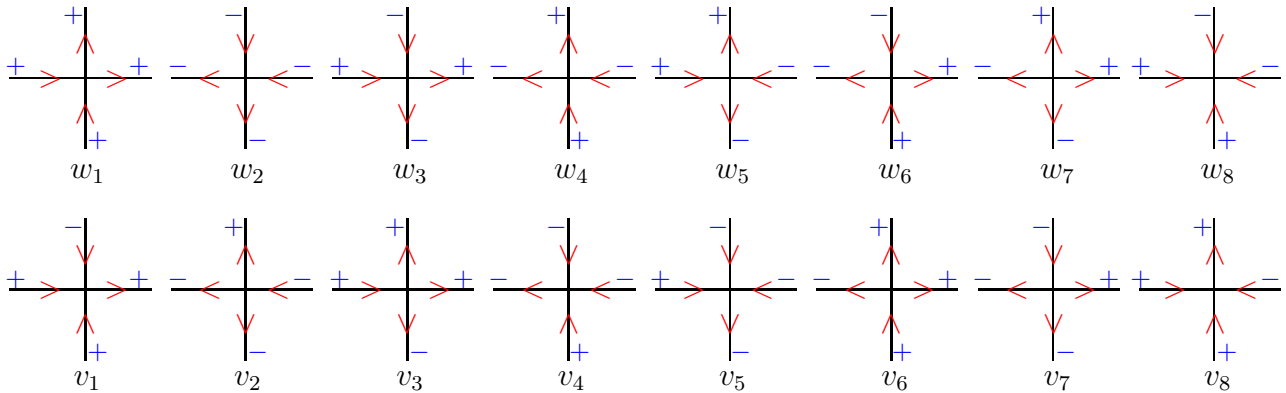

We observe that
weights can be organized in terms of two distinct families 
of eight vertex states according to the even 
or odd number arrows orientations at a vertex. 
The standard eight vertex model \cite{BAX}
corresponds to the system in which the number of in and out arrows are even 
and the respective weights are denoted by $w_1,\dots,w_8$. The vertex model with an 
odd number of in and out arrows 
has been denominated odd eight-vertex model which encompasses the classical 
Ashkin-Teller model \cite{WKodd}. The corresponding
vertex weights are indicated here by $v_1,\dots,v_8$. 
We remark that our notation for the even vertex weights is the most common choice in the
literature \cite{LIWU} while the  representation of the odd vertex weights 
is done by reversing the up vertical states.
The partition function of the full sixteen-vertex model can be defined
by the following sum,
\begin{equation}
\label{PAR16V}
Z_{16v}(L)= \sum_{arrows} \left( w_1^{n_1} \dots w_8^{n_8} \right) \left( v_{1}^{m_1} \dots v_8^{m_8} \right)
\end{equation}
where the summation is extended over all possible arrow configurations, the integers $n_i$ and $m_j$ indicate 
respectively the
total number of weights $w_i$ and $v_j$ in given lattice configuration. We assume periodic boundary 
conditions for  arrows
configurations in both lattice directions.

Over the years certain mapping relations between 
two state spin models
and the eight-vertex and the sixteen-vertex 
models 
on the square lattice have been introduced 
in the literature. There is a mapping 
between the Ising model with zero magnetic field and the eight-vertex
model with even weights in which the spin variables lie on the
faces of the square lattice \cite{FWU,WU,KAD}. This is however a 
two-to-one correspondence and the partition function of the Ising model
turns out to be twice of that of the equivalent even eight-vertex model.
In another mapping  one
introduces the spin variables on the medial 
points of the square
lattice and the configurations 
for the four spins around a vertex give rises to 
sixteen-vertex weights \cite{SUFI}. By using this
equivalence 
the Ising model 
in the presence of a magnetic field can be mapped 
to a specific sixteen-vertex model
on the square lattice \cite{LIWU,GAA}. We observe
that in such correspondence
the spin model
has twice as many lattice points as the vertex 
model and 
in the thermodynamic limit the free-energy of
the Ising model in a magnetic field is half of that 
of the equivalent
sixteen-vertex model. However, there is a more direct 
formulation
of the isotropic Ising model 
with non-zero magnetic field as a sixteen-vertex 
problem which 
uses as intermediate
step a mapping to a lattice gas on the 
square lattice \cite{LIWU}.
The interesting feature
of this equivalence is that it is valid for
toroidal square lattice with a finite number 
of lattice points. 
We remark that this type of correspondence
has been further elaborated in \cite{SMA} yet 
restricted to the
case of 
isotropic $J_v=J_h$ interactions. For later comparison with our 
results we present 
below the expressions of the weights of the equivalent 
sixteen-vertex model. It turns out that
the partition functions of the isotropic Ising model in a magnetic field (\ref{PARISI})
and of the sixteen-vertex model (\ref{PAR16V})  coincide
provided that the vertex
weights are given by \cite{LIWU},
\begin{eqnarray}
&& \omega_1= 2 \cosh(\beta H) [\cosh(\beta J_h)]^2,~~~\omega_2=2\cosh(\beta H) [\sinh(\beta J_h)]^2, \nonumber \\
&& \omega_3=\omega_4=\omega_5=\omega_6=\omega_7=\omega_8= \cosh(\beta H) \sinh(2\beta J_h), \nonumber \\
&& v_1=v_3=v_6=v_8= 2\sinh(\beta H) [\cosh(\beta J_h)]^2 \sqrt{\tanh(\beta J_h)}, \nonumber \\
&& v_2=v_4=v_5=v_7= \frac{2\sinh(\beta H) [\sinh(\beta J_h)]^2} {\sqrt{\tanh(\beta J_h)}}
\end{eqnarray}

We next mention that equivalences between spin and vertex models with arbitrary 
number of local states
have also been pursued in the literature \cite{PER,GUT}. 
These correspondences involve somehow either
a different number of states for the spin and vertex models or if the
number of the states are equal the respective mapping is of
multiplicity two. One of the purpose of 
this paper is to associate to any $n$-state spin model with next-neighbor 
interactions an equivalent
$n$-state vertex model both defined on the square lattice. Our mapping is in 
the sense 
of the last mentioned equivalence among the isotropic Ising model in a nonzero 
magnetic field and the
sixteen-vertex model. This means that our correspondence implies that the 
partition functions 
of the $n$-state spin and vertex models coincide for finite size $L$ with
toroidal boundary conditions. It turns out that the equivalent $n$-state 
vertex model has only $n^3$ non-null weights and therefore our mapping 
is more economical than previous equivalences established 
for the Ising model in the presence of 
a magnetic field. In fact, instead of 
a sixteen-vertex model
our results leads us to consider a mixed eight-vertex model 
in which four weights 
have an even number of arrows while the other four weights have 
an odd number
of arrows at the vertex. More precisely, the partition function of such
equivalent mixed eight-vertex model is defined as,
\begin{equation}
\label{PARM8V}
Z_{m8v}(L)= \sum_{arrows} \left(w_1^{n_1} w_2^{n_2} w_5^{n_5} w_6^{n_6} \right) 
\left( v_1^{m_1} v_2^{m_2} v_5^{m_5} v_6^{m_6} \right) 
\end{equation}
where the summation is similar to the one already 
defined for the general sixteen vertex model.

In this paper we  argue that the partition functions 
of the Ising model 
in a magnetic field (\ref{PARISI}) and that of the 
mixed eight-vertex model (\ref{PARM8V}) are the same for
particular choices of the vertex weights. The dependence of the 
vertex weights with
the Ising model edge interactions is
exhibited in Fig.(\ref{FigMap}). We emphasize that such relationship is valid for 
an anisotropic Ising model with generic 
couplings $J_h$ and $J_v$
in the presence of an external magnetic field.
\setlength{\unitlength}{2500sp}
\begin{figure}[ht]
\begin{center}
\begin{picture}(8000,2000)
{\put(-2200,900){\line(1,0){1500}}}
{\put(1200,900){\line(1,0){1500}}}
{\put(4400,900){\line(1,0){1500}}}
{\put(7600,900){\line(1,0){1500}}}
{\put(-2200,-1200){\line(1,0){1500}}}
{\put(1200,-1200){\line(1,0){1500}}}
{\put(4400,-1200){\line(1,0){1500}}}
{\put(7600,-1200){\line(1,0){1500}}}
{\put(-1450,1600){\line(0,-1){1500}}}
{\put(1980,1600){\line(0,-1){1500}}}
{\put(5200,1600){\line(0,-1){1500}}}
{\put(8400,1600){\line(0,-1){1500}}}
{\put(-1450,-500){\line(0,-1){1500}}}
{\put(1980,-500){\line(0,-1){1500}}}
{\put(5200,-500){\line(0,-1){1500}}}
{\put(8400,-500){\line(0,-1){1500}}}
{\put(-1800,890){\makebox(0,0){\fontsize{12}{14}\selectfont $\color{red} >$}}}
{\put(1665,890){\makebox(0,0){\fontsize{12}{14}\selectfont $\color{red} <$}}}
{\put(4835,890){\makebox(0,0){\fontsize{12}{14}\selectfont $\color{red} >$}}}
{\put(8045,890){\makebox(0,0){\fontsize{12}{14}\selectfont $\color{red} <$}}}
{\put(-2140,1010){\makebox(0,0){\fontsize{10}{12}\selectfont $\color{blue} +$}}}
{\put(1270,1010){\makebox(0,0){\fontsize{10}{12}\selectfont $\color{blue} -$}}}
{\put(4480,1010){\makebox(0,0){\fontsize{10}{12}\selectfont $\color{blue} +$}}}
{\put(7670,1010){\makebox(0,0){\fontsize{10}{12}\selectfont $\color{blue} -$}}}
{\put(-1800,-1200){\makebox(0,0){\fontsize{12}{14}\selectfont $\color{red} >$}}}
{\put(1665,-1200){\makebox(0,0){\fontsize{12}{14}\selectfont $\color{red} <$}}}
{\put(4835,-1200){\makebox(0,0){\fontsize{12}{14}\selectfont $\color{red} >$}}}
{\put(8045,-1200){\makebox(0,0){\fontsize{12}{14}\selectfont $\color{red} <$}}}
{\put(-2140,-1080){\makebox(0,0){\fontsize{10}{12}\selectfont $\color{blue} +$}}}
{\put(1270,-1080){\makebox(0,0){\fontsize{10}{12}\selectfont $\color{blue} -$}}}
{\put(4480,-1080){\makebox(0,0){\fontsize{10}{12}\selectfont $\color{blue} +$}}}
{\put(7670,-1080){\makebox(0,0){\fontsize{10}{12}\selectfont $\color{blue} -$}}}
{\put(-1140,890){\makebox(0,0){\fontsize{12}{14}\selectfont $\color{red} >$}}}
{\put(2320,890){\makebox(0,0){\fontsize{12}{14}\selectfont $\color{red} <$}}}
{\put(5560,890){\makebox(0,0){\fontsize{12}{14}\selectfont $\color{red} <$}}}
{\put(8720,890){\makebox(0,0){\fontsize{12}{14}\selectfont $\color{red} >$}}}
{\put(-800,1010){\makebox(0,0){\fontsize{10}{12}\selectfont $\color{blue} +$}}}
{\put(2590,1010){\makebox(0,0){\fontsize{10}{12}\selectfont $\color{blue} -$}}}
{\put(5800,1010){\makebox(0,0){\fontsize{10}{12}\selectfont $\color{blue} -$}}}
{\put(9000,1010){\makebox(0,0){\fontsize{10}{12}\selectfont $\color{blue} +$}}}
{\put(-1140,-1200){\makebox(0,0){\fontsize{12}{14}\selectfont $\color{red} >$}}}
{\put(2320,-1200){\makebox(0,0){\fontsize{12}{14}\selectfont $\color{red} <$}}}
{\put(5560,-1200){\makebox(0,0){\fontsize{12}{14}\selectfont $\color{red} <$}}}
{\put(8720,-1200){\makebox(0,0){\fontsize{12}{14}\selectfont $\color{red} >$}}}
{\put(-800,-1080){\makebox(0,0){\fontsize{10}{12}\selectfont $\color{blue} +$}}}
{\put(2590,-1080){\makebox(0,0){\fontsize{10}{12}\selectfont $\color{blue} -$}}}
{\put(5800,-1080){\makebox(0,0){\fontsize{10}{12}\selectfont $\color{blue} -$}}}
{\put(9000,-1080){\makebox(0,0){\fontsize{10}{12}\selectfont $\color{blue} +$}}}
{\put(-1450,1240){\makebox(0,0){\fontsize{12}{14}\selectfont $\color{red} \wedge$}}}
{\put(1980,1240){\makebox(0,0){\fontsize{12}{14}\selectfont $\color{red} \vee$}}}
{\put(5200,1240){\makebox(0,0){\fontsize{12}{14}\selectfont $\color{red} \wedge$}}}
{\put(8400,1240){\makebox(0,0){\fontsize{12}{14}\selectfont $\color{red} \vee$}}}
{\put(-1570,1520){\makebox(0,0){\fontsize{10}{12}\selectfont $\color{blue} +$}}}
{\put(1860,1520){\makebox(0,0){\fontsize{10}{12}\selectfont $\color{blue} -$}}}
{\put(5060,1520){\makebox(0,0){\fontsize{10}{12}\selectfont $\color{blue} +$}}}
{\put(8260,1520){\makebox(0,0){\fontsize{10}{12}\selectfont $\color{blue} -$}}}
{\put(-1450,-860){\makebox(0,0){\fontsize{12}{14}\selectfont $\color{red} \vee$}}}
{\put(1980,-860){\makebox(0,0){\fontsize{12}{14}\selectfont $\color{red} \wedge$}}}
{\put(5200,-860){\makebox(0,0){\fontsize{12}{14}\selectfont $\color{red} \vee$}}}
{\put(8400,-860){\makebox(0,0){\fontsize{12}{14}\selectfont $\color{red} \wedge$}}}
{\put(-1570,-580){\makebox(0,0){\fontsize{10}{12}\selectfont $\color{blue} -$}}}
{\put(1860,-580){\makebox(0,0){\fontsize{10}{12}\selectfont $\color{blue} +$}}}
{\put(5060,-580){\makebox(0,0){\fontsize{10}{12}\selectfont $\color{blue} -$}}}
{\put(8260,-580){\makebox(0,0){\fontsize{10}{12}\selectfont $\color{blue} +$}}}
{\put(-1450,450){\makebox(0,0){\fontsize{12}{14}\selectfont $\color{red} \wedge$}}}
{\put(1980,450){\makebox(0,0){\fontsize{12}{14}\selectfont $\color{red} \vee$}}}
{\put(5200,450){\makebox(0,0){\fontsize{12}{14}\selectfont $\color{red} \vee$}}}
{\put(8400,450){\makebox(0,0){\fontsize{12}{14}\selectfont $\color{red} \wedge$}}}
{\put(-1330,180){\makebox(0,0){\fontsize{10}{12}\selectfont $\color{blue} +$}}}
{\put(2120,180){\makebox(0,0){\fontsize{10}{12}\selectfont $\color{blue} -$}}}
{\put(5340,180){\makebox(0,0){\fontsize{10}{12}\selectfont $\color{blue} -$}}}
{\put(8540,180){\makebox(0,0){\fontsize{10}{12}\selectfont $\color{blue} +$}}}
{\put(-1450,-1640){\makebox(0,0){\fontsize{12}{14}\selectfont $\color{red} \wedge$}}}
{\put(1980,-1640){\makebox(0,0){\fontsize{12}{14}\selectfont $\color{red} \vee$}}}
{\put(5200,-1640){\makebox(0,0){\fontsize{12}{14}\selectfont $\color{red} \vee$}}}
{\put(8400,-1640){\makebox(0,0){\fontsize{12}{14}\selectfont $\color{red} \wedge$}}}
{\put(-1330,-1920){\makebox(0,0){\fontsize{10}{12}\selectfont $\color{blue} +$}}}
{\put(2120,-1920){\makebox(0,0){\fontsize{10}{12}\selectfont $\color{blue} -$}}}
{\put(5340,-1920){\makebox(0,0){\fontsize{10}{12}\selectfont $\color{blue} -$}}}
{\put(8540,-1920){\makebox(0,0){\fontsize{10}{12}\selectfont $\color{blue} +$}}}
{\put(-1430,-140){\makebox(0,0){\fontsize{12}{14}\selectfont $e^{\beta(J_h+J_v +H)}$}}}
{\put(2020,-140){\makebox(0,0){\fontsize{12}{14}\selectfont $e^{\beta(J_h+J_v -H)}$}}}
{\put(5220,-140){\makebox(0,0){\fontsize{12}{14}\selectfont $e^{\beta(-J_h+J_v +\frac{H}{2})}$}}}
{\put(8420,-140){\makebox(0,0){\fontsize{12}{14}\selectfont $e^{\beta(-J_h+J_v -\frac{H}{2})}$}}}
{\put(-1430,-2220){\makebox(0,0){\fontsize{12}{14}\selectfont $e^{\beta(J_h-J_v +\frac{H}{2})}$}}}
{\put(2020,-2220){\makebox(0,0){\fontsize{12}{14}\selectfont $e^{\beta(J_h-J_v -\frac{H}{2})}$}}}
{\put(5220,-2220){\makebox(0,0){\fontsize{12}{14}\selectfont $e^{\beta(-J_h-J_v)}$}}}
{\put(8420,-2220){\makebox(0,0){\fontsize{12}{14}\selectfont $e^{\beta(-J_h-J_v)}$}}}
\end{picture}
\end{center}
\vspace{2.3cm}
\caption{The equivalence of the Ising model in a magnetic field 
with a mixed 
eight-vertex model on the square lattice
with toroidal boundary conditions.}
\label{FigMap}
\end{figure}
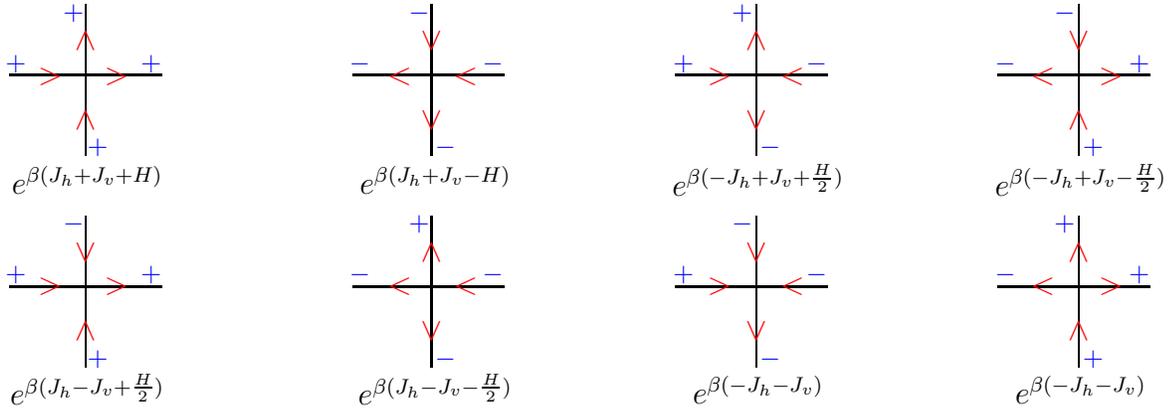

We have organized this paper as follows. In next section we 
introduce 
the $n$-state spin and vertex
models and formulate their partition functions in terms of 
the transfer matrix concept. In section 3 we describe 
two possible correspondences of the
among the $n$-state spin model and a $n$-state 
vertex model with $n^3$ non-null weights. We use 
the Hamiltonian 
limit to built an anstaz for the main
structure of the
weights of the equivalent $n$-state vertex model. The match of the 
partition functions 
is done by comparing
the respective transfer matrices operators for $n \leq 4$ 
and $L \leq 6$ and we conjecture that our mappings
should be valid 
for general $n$ and $L$.
In section 4 we study the Yang-Baxter 
algebra for the mixed eight-vertex model on the subspace 
of symmetrical weights. 
This manifold encodes the Ising
model without an external magnetic field. We find a 
solution to the 
Yang-Baxter relation in
which the respective $\mathrm{R}$-matrix is not 
expressible in terms 
of the difference of the spectral variables 
parameterizing the respective Lax operator. The corresponding spin chain
is shown to be related to that of the $XY$ model 
in a transverse magnetic field with a Dzyaloshinky-Moriya interaction.
This has motived us in section 5 to investigate possible mapping among
the mixed eight-vertex model and the even eight-vertex model with weights
satisfying the free-fermion condition. As a byproduct of this analysis we
present novel correspondences among the  
Ising model in absence of magnetic field
and the free-fermion even eight-vertex model such that 
their partition functions are exactly the same 
for toroidal finite lattice. In section 6 we discuss the possibility of embedding an integrable
spin model on the context of the quantum inverse scattering framework. In particular, we have
determined the underlying $\mathrm{R}$-matrix of the 27-vertex model equivalent to the $N=3$ Fateev-Zamolodchikov
spin model.
In the appendices we summarize some of
technical details we have omitted in the main text.

\section{The $n$-state spin and vertex models}

The $n$-state spin 
lattice model with nearest neighbors interactions can be built out of
$n^2$ horizontal and $n^2$ vertical 
edge interactions weights \cite{BAX,PER1}. Let us denote 
by $\sigma_{i,j}$ the state variables at the site $(i,j)$ 
of the square lattice of size $L$. We then associate local
horizontal  
$W_{h}(\sigma_{i,j},\sigma_{i,j+1})$ and vertical 
$W_{v}(\sigma_{i,j},\sigma_{i+1,j})$ Boltzmann weights to
characterize the energy interactions 
among two neighboring spins. 
These edge weights are schematically shown in Fig.(\ref{FigSpin}).
\setlength{\unitlength}{2500sp}
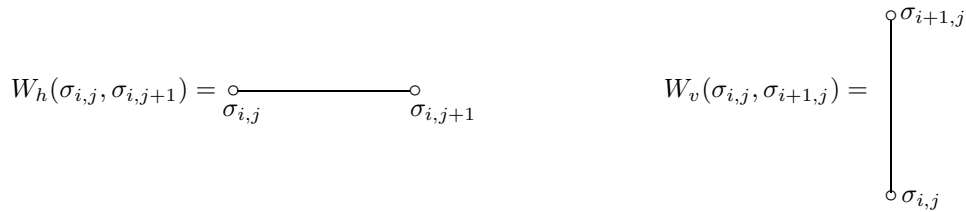
\begin{figure}[ht]
\begin{center}
\begin{picture}(8000,2000)
{\put(960,900){\line(1,0){1700}}}
{\put(920,900){\makebox(0,0){\fontsize{10}{10}\selectfont $\circ$}}}
{\put(1010,670){\makebox(0,0){\fontsize{10}{10}\selectfont $\sigma_{i,j}$}}}
{\put(2720,900){\makebox(0,0){\fontsize{10}{10}\selectfont $\circ$}}}
{\put(3000,670){\makebox(0,0){\fontsize{10}{10}\selectfont $\sigma_{i,j+1}$}}}
{\put(-270,900){\makebox(0,0){\fontsize{10}{10}\selectfont $W_{h}(\sigma_{i,j},\sigma_{i,j+1})=$}}}
{\put(7440,1600){\line(0,-1){1700}}}
{\put(7440,1650){\makebox(0,0){\fontsize{10}{10}\selectfont $\circ$}}}
{\put(7440,-140){\makebox(0,0){\fontsize{10}{10}\selectfont $\circ$}}}
{\put(7740,-180){\makebox(0,0){\fontsize{10}{10}\selectfont $\sigma_{i,j}$}}}
{\put(7860,1640){\makebox(0,0){\fontsize{10}{10}\selectfont $\sigma_{i+1,j}$}}}
{\put(6200,900){\makebox(0,0){\fontsize{10}{10}\selectfont $W_{v}(\sigma_{i,j},\sigma_{i+1,j})=$}}}
\end{picture}
\end{center}
\caption{The horizontal 
$W_{h}(\sigma_{i,j},\sigma_{i,j+1})$
and the vertical 
$W_{v}(\sigma_{i,j},\sigma_{i+1,j})$
local Boltzmann weights of spin models. The spin variables $\sigma_{i,j}$ can 
take $n$ possible values.}
\label{FigSpin}
\end{figure}

The respective partition function is the sum of the product of all 
the local Boltzmann weights
which can be written as follows,
\begin{equation}
Z_{\mathrm{spin}}(L)= \sum_{<\sigma_{i,j}>} \prod_{i,j=1}^{L} W_h\left(\sigma_{i,j},\sigma_{i,j+1}\right)
W_v\left(\sigma_{i,j},\sigma_{i+1,j}\right)
\label{PARSPIN}
\end{equation}
where sum is over all allowed spin $\sigma_{i,j}$ configurations 
on the lattice. 
Periodic boundary
conditions are imposed considering the 
identifications $\sigma_{i,L+1}=\sigma_{i,1}$ and 
$\sigma_{L+1,j}=\sigma_{1,j}$. 

It is well known that the partition 
function (\ref{PARSPIN}) 
can be obtained as the trace 
of successive matrix multiplications
of an operator called transfer matrix \cite{KAM}. For the spin model it is 
sometimes convenient to
consider the diagonal-to-diagonal 
transfer matrix $T_{\mathrm{dia}}(L)$ which is built
on the lattice states along the diagonals \cite{ONSA,MIT}. We recall that this operator plays an 
important role on the construction
of integrable spin models derived from a family of commuting transfer matrices \cite{BAX,PER1}. 
Considering periodic 
conditions
in the horizontal the elements of the diagonal-to-diagonal transfer matrix 
are,
\begin{equation}
\label{TDIA}
\left[T_{\mathrm{dia}}(L)\right]_{a_1,\dots,a_L}^{b_1,\dots,b_L}= \prod_{j=1}^{L} W_v(a_j,b_j) W_h(a_j,b_{j+1})
\end{equation}
where $b_{L+1}=b_1$. We observe that the transfer matrix 
is defined in the so-called quantum space  
${\cal V}= \displaystyle \prod_{j=1}^{L} \otimes {\mathrm{C}}^{n}$ and the partition function  can be obtained as follows,
\begin{equation}
\label{Zspin}
Z_{\mathrm{spin}}(L)= \mathrm{Tr}_{{\cal V}} \left[ T_{\mathrm{dia}}(L) \right]^{L}
\end{equation}

In the vertex models the local configurations are defined by the
spin variables attached to the four links of the square lattice 
joining together at the vertex \cite{BAX,PER1}. To a given 
vertex at site $(i,j)$ we associate a Boltzmann weight 
$w\left(\alpha_{ij},\alpha_{ij+1}|\gamma_{ij},\gamma_{i+1j}\right)$ 
as exhibited 
in Fig.(\ref{FigVer}). Here we assume that the horizontal and 
vertical spin
variables $\alpha_{ij}$ and $\gamma_{ij}$ take values on the
same finite set constituted of $n$ states. This means that the total 
number of weights 
defining this
vertex model is therefore $n^4$.
\setlength{\unitlength}{2500sp}
\begin{figure}[ht]
\begin{center}
\begin{picture}(8000,2000)
{\put(3560,900){\line(1,0){1900}}}
{\put(4500,1800){\line(0,-1){1900}}}
{\put(3280,900){\makebox(0,0){\fontsize{10}{10}\selectfont $\gamma_{ij}$}}}
{\put(5800,900){\makebox(0,0){\fontsize{10}{10}\selectfont $\gamma_{i+1j}$}}}
{\put(4500,-200){\makebox(0,0){\fontsize{10}{10}\selectfont $\alpha_{ij}$}}}
{\put(4500,1960){\makebox(0,0){\fontsize{10}{10}\selectfont $\alpha_{ij+1}$}}}
{\put(900,900){\makebox(0,0){\fontsize{10}{10}\selectfont $w\left(\alpha_{ij},\alpha_{ij+1}|\gamma_{ij},\gamma_{i+1j}\right)=$}}}
\end{picture}
\end{center}
\caption{The 
local Boltzmann weights of vertex models at the $(i,j)$ lattice site.
Both spin variables $\alpha_{ij}$ and $\gamma_{ij}$ can take $n$ possible values.}
\label{FigVer}
\end{figure}
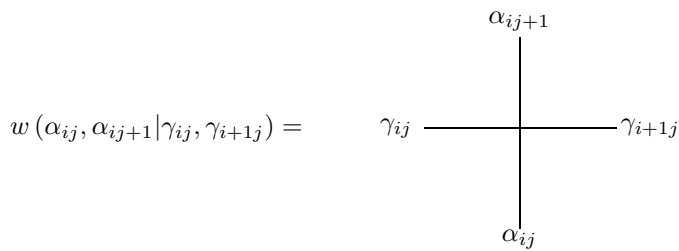

The partition function associated to the 
such $n$-state vertex models 
on the square lattice 
can be written by the following expression,
\begin{equation}
Z_{\mathrm{ver}}(L)= \sum_{<\alpha_{ij},\gamma_{ij}>} \prod_{i,j=1}^{L} 
w\left(\alpha_{ij},\alpha_{ij+1}|\gamma_{ij},\gamma_{i+1j}\right) 
\end{equation}
where sum is over all allowed horizontal $\alpha_{ij}$ and 
vertical $\gamma_{ij}$ spin configurations 
on the lattice. 
Periodic boundary
conditions are imposed considering the 
identifications $\alpha_{iL+1}=\alpha_{i1}$ and 
$\gamma_{L+1j}=\gamma_{1j}$.

The vertex models has the advantage of having an underlying 
tensor structure. This
plays an important role in connection with quantum spin chains and to the
formulation of a quantum  version of the
inverse scattering method \cite{FAD}.
This property allows one to represent the 
partition function under toroidal boundary conditions as the
trace over two spaces associated to the horizontal 
and vertical
degrees of freedom of the vertex weights. To this end we define 
a set of matrices named Lax operators as follows,
\begin{equation}
\label{LAX}
{\mathbb{L}}_{{\cal{A}}j}= \sum_{i_1,i_2,i_3,i_4=1}^{n} w(i_1,i_2|i_3,i_4) e^{(j)}_{i_1,i_2} \otimes e_{i_3,i_4},~~~j=1,\dots,L
\end{equation}
where $e_{i_1,i_2}$ denotes the $n \times n$ matrix with only one 
non-vanishing entry with value 1 at row
$i_1$ and column $i_2$. This is the basis of the space of the 
Lax operator denominated 
auxiliary or horizontal space
${\cal{A}}={\mathrm C}^n$. The vertical degrees of freedom gives rise 
to the quantum space basis 
$e_{i_1,i_2}^{(j)} \in {\cal V} $ defined as,
\begin{equation}
e_{i_1,i_2}^{(j)}=\prod_{\stackrel{k=1}{k \neq j}}^{L} \mathrm{I}_n^{(k-1)} \otimes  e_{i_1,i_2} \otimes  \mathrm{I_n}^{(L-k)} 
\end{equation}
where
$\mathrm{I_n}$ is the $n \times n$ identity matrix.

By virtue of the periodic boundary condition on the horizontal direction 
the row-to-row transfer 
matrix $T_{\mathrm{ver}}(L)$ associated 
to the vertex models can be 
written in a compact form as the trace over the auxiliary space of
an ordered product of Lax operators,
\begin{equation}
\label{TVER}
T_{\mathrm{ver}}(L)= Tr_{{\cal{A}}}\left[ {\mathbb{L}}_{{\cal A} L}  
{\mathbb{L}}_{{\cal A} L-1} \dots  
{\mathbb{L}}_{{\cal A} 1 } \right] 
\end{equation}
where $T_{\mathrm{ver}}(L)$ is again an operator belonging
to the quantum space ${\cal V}= \displaystyle \prod_{j=1}^{L} \otimes {\mathrm{C}}^{n}$. 

Once again the 
partition function of the vertex model can be obtained 
by multiplying layers of transfer
matrices and for toroidal boundary conditions we obtain,
\begin{equation}
\label{Zvertex}
Z_{\mathrm{ver}}(L)= \mathrm{Tr}_{{\cal V}}\left[ T_{\mathrm{ver}}(L)\right ]^{L}
\end{equation}

In next section we shall argue that the partitions functions $Z_{\mathrm{spin}}(L)$ and 
$Z_{\mathrm{ver}}(L)$ coincide
for a suitable choice of the vertex weights $w(i_1,i_2|i_3,i_4)$.

\section{The spin-vertex correspondence}

We start by noticing that the edge weights of the $n$-state spin 
model can be seen as coordinates of the product of two projective
spaces while the 
Boltzmann weights of a $n$-state vertex model may be interpreted 
as points of a single projective space.
Here we would like to discuss a map $\varphi$ among these two projective spaces 
of the form, 
\begin{equation}
\label{blow}
\renewcommand{\arraystretch}{1.5}
\begin{array}{ccc}
P^{n^2-1} \times P^{n^2-1} & \overset{\varphi}{\longrightarrow} &  P^{m} \\
W_h(i_1,i_2), W_v(i_3,i_4) & \longmapsto & w(i_1,i_2|i_3,i_4)
\end{array}
\end{equation}
where $m \leq n^4-1$ since some of the vertex weights may be zero.

In order to shed some light on the structure 
of the 
vertex weights we investigate the Hamiltonian 
limit of both spin and vertex models. For the spin model the underlying spin chain Hamiltonian
is obtained by expanding the edge weights around a point in which the diagonal-to-diagonal transfer
matrix reduces to the identity matrix, see for instance \cite{ALLI,MCPER1,MCPER2}. To this end 
we start our
analysis by assuming 
that the edge weights 
can be expanded 
in terms of some
spectral parameter denoted here by 
$\varepsilon$. We next consider that 
the expansion of 
the edge weights around $\varepsilon=0$ up to the
first order is given by,
\begin{equation}
W_h(i_1,i_2;\varepsilon) \sim  1 +\varepsilon {{\dot{W}}}_h(i_1,i_2)+\mathcal{O}(\varepsilon^2),~~ 
W_v(i_1,i_2;\varepsilon) \sim  \delta_{i_1,i_2} +\varepsilon {{\dot{W}}}_v(i_1,i_2) 
+\mathcal{O}(\varepsilon^2)
\end{equation}
where ${{\dot{W}}}_h(i_1,i_2)$ and 
${{\dot{W}}}_v(i_1,i_2)$ denote the edge weights expansion
coefficients. 

At this point we note that at $\varepsilon=0$ the diagonal-to-diagonal
transfer matrix (\ref{TDIA}) indeed becomes 
the identity matrix $I_{d}$ with dimension $d=n^L$. The Hamiltonian limit is 
obtained by expanding
the transfer matrix (\ref{TDIA}) about the parameter 
$\varepsilon$ and up to the first order we have,
\begin{equation}
T_{\mathrm{dia}}(L;\varepsilon) \sim I_{d} +\varepsilon \left[ \sum_{j=1}^{L-1} H^{\mathrm{(spin)}}_{j,j+1} + H^{\mathrm{(spin)}}_{L,1}\right]
\end{equation}
where the two-body Hamiltonian 
$H^{\mathrm{(spin)}}_{j,j+1}$ is given by,
\begin{equation}
\label{TWOSP}
H^{\mathrm{(spin)}}_{j,j+1}= 
\sum_{i_1,i_2=1}^{n} {{\dot{W}}}_h(i_1,i_2) e_{i_1,i_1}^{(j)} \otimes e_{i_2,i_2}^{(j+1)} +
\sum_{i_1,i_2,i_3=1}^{n} {{\dot{W}}}_v(i_1,i_2) e_{i_1,i_2}^{(j)} \otimes e_{i_3,i_3}^{(j+1)}
\end{equation}

For the vertex model the Hamiltonian limit 
is considered by expanding 
the logarithm of the
row-to-row transfer matrix (\ref{TVER}) around a point 
in which the respective Lax operator reduces to the
permutator defined on the tensor product 
${\mathrm{C}}^{n}
\otimes {\mathrm{C}}^{n}$ \cite{BAXH}. We next 
assume that the expansion of the vertex weights
giving the permutator at zero order is as follows,
\begin{equation}
w(i_1,i_2|i_3,i_4;\varepsilon) \sim \delta_{i_1,i_4} \delta_{i_2,i_3} +\varepsilon {{\dot{w}}}(i_1,i_2|i_3,i_4) 
+\mathcal{O}(\varepsilon^2) 
\end{equation}
where ${{\dot{w}}}(i_1,i_2|i_3,i_4)$ is the vertex weights 
first order expansion coefficients. By expanding the logarithm 
of the transfer 
matrix about $\varepsilon$ we obtain 
to the first order, 
\begin{equation}
T^{-1}_{\mathrm{ver}}(L;\varepsilon=0)
T_{\mathrm{ver}}(L;\varepsilon) \sim I_{d} +\varepsilon \left[ \sum_{j=1}^{L-1} H^{\mathrm{(ver)}}_{j,j+1} + H^{\mathrm{(ver)}}_{L,1}\right]
\end{equation}
where the two-body Hamiltonian 
$H^{\mathrm{(ver)}}_{j,j+1}$ is,
\begin{equation}
\label{TWOVER}
H^{\mathrm{(ver)}}_{j,j+1}= 
\sum_{i_1,i_2,i_3,i_4=1}^{n}  
{{\dot{w}}}(i_1,i_2|i_3,i_4) 
e_{i_3,i_2}^{(j)} \otimes e_{i_1,i_4}^{(j+1)}
\end{equation}

The next step is to compare the expressions for the spin and vertex
two-body Hamiltonians (\ref{TWOSP},\ref{TWOVER}). We find that they
can be matched up to the first order in $\varepsilon$ once we have
the following relationships,
\begin{eqnarray}
w(i_1,i_2|i_3,i_4,\varepsilon) & \sim & \delta_{i_2,i_3} \delta_{i_1,i_4} +\varepsilon \left (
\delta_{i_2,i_3} {{\dot{W}}}_h(i_3,i_1)+ 
{{\dot{W}}}_v(i_3,i_2) \right) \delta_{i_1,i_4} \nonumber \\
& \sim & \left(1+\varepsilon
{{\dot{W}}}_h(i_3,i_1) \right)
\left (\delta_{i_3,i_2} +\varepsilon 
{{\dot{W}}}_v(i_3,i_2) \right) \delta_{i_1,i_4} \nonumber \\
& \sim &
W_h(i_3,i_1;\varepsilon) W_v(i_3,i_2;\varepsilon) \delta_{i_1,i_4}
\end{eqnarray}

Motivated by the last identification we propose 
our ansatz for
the mapping among the edge and vertex weights, namely
\begin{equation}
\label{ansa}
w(i_1,i_2|i_3,i_4)= W_h(i_3,i_1) W_v(i_3,i_2) \delta_{i_1,i_4}
\end{equation}
and therefore the equivalent vertex model has $n^3$ non-null weights.

We now start presenting our evidences that the 
ansatz (\ref{ansa}) 
should imply that the partition functions
of the spin and vertex models coincide. To this end we turn our
attention to the computation of the vertex model row-to-row transfer
matrix (\ref{TVER}) for some values of $n$ and $L$. For $L=1$ the
transfer matrix is just sum of matrices associated to the  
diagonal partitions of
the Lax operators. By using the mapping relation (\ref{ansa}) 
its matrix elements can be computed for arbitrary $n$ to be,
\begin{equation}
\left[T_{\mathrm{ver}}(L=1)\right]_{a_1}^{b_1}= W_h(a_1,b_1) W_v(a_1,b_1)
\end{equation}

For $L=2$ we have to compute the product of partitions of 
two Lax operators and this
calculation is in general  involved for 
an arbitrary $n$-state vertex model. In our case however the
equivalent vertex model has a number of suitable null weights 
and we have find the matrices elements of such products 
of partitions
are single monomials constituted by the product 
of two vertex weights. We have carried
out these computations explicitly for $ n \leq 4$ and after 
using the proposal (\ref{ansa}) we found
that the matrix elements of the row-row transfer matrix
can be organized as follows,
\begin{equation}
\left[T_{\mathrm{ver}}(L=2)\right]_{a_1,a_2}^{b_1,b_2}= W_h(a_1,b_1) W_v(a_1,b_2) W_h(a_2,b_2) W_v(a_2,b_1)
\end{equation}

We observe that results for $L=1,2$ have indeed a simple 
pattern very similar to that of 
diagonal-to-diagonal transfer matrix, see Eq.(\ref{TDIA}). This fact permits
us to guess what could be the structure
of the matrix elements of the row-to-row transfer 
matrix for arbitrary $L$, namely
\begin{equation}
\label{Trowdia}
\left[T_{\mathrm{ver}}(L)\right]_{a_1,\dots,a_L}^{b_1,\dots,b_L}= \prod_{j=1}^{L} W_h(a_j,b_j) W_v(a_j,b_{j+1})
\end{equation}
where $b_{L+1}=b_1$. 
With the help of symbolic algebra packages we have verified 
for $n \leq 4$ and $L \leq 6$ that the formulae (\ref{Trowdia})
indeed produce the matrix elements of the row-to-row 
transfer matrix (\ref{TVER}) with the vertex weights (\ref{ansa}). 
In the case of two-state models we managed to check this result up to $L=10$.
Based on this checking we conjecture that
such result should be valid for arbitrary $n$ and $L$ but a systematic proof of that has eluded us so far.

We finally note that the matrices elements (\ref{Trowdia}) is same 
of the diagonal-to-diagonal
transfer matrix (\ref{TDIA}) by interchanging 
the horizontal and vertical edge weights. 
This is equivalent to rotation of the lattice by $90^{0}$ degrees and
certainly the partition function 
is invariant under such transformation.
Therefore, under the assumption 
of the validity 
of the expression (\ref{Trowdia}) for arbitrary $n$ and $L$ we can 
write
the partition function of the spin model as,
\begin{equation}
Z_{\mathrm{spin}}(L)= \mathrm{Tr}_{{\cal V}}\left[ T_{\mathrm{ver}}(L)\right ]^{L}
\end{equation}
provided that the vertex and the edge 
weights are related by Eq.(\ref{ansa}). 

This leads us to conjecture that the partition functions 
of the spin and vertex models 
coincide when the
respective Boltzmann weights satisfy the relation 
given by Eq.(\ref{ansa}). We stress that this result
does not depend on the fact that the 
spin and vertex models 
have necessarily an underlying Hamiltonian limit.

\subsection{Another equivalence}

In the transfer matrix method we have first to define the respective 
layers of the system. As such we
can choose the layers based on the configurations of rows of spins 
as originally devised for the Ising model \cite{KAM,KOGU}. This gives 
rise to the 
row-to-row transfer matrix 
$T_{\mathrm{row}}(L)$ in
which we can separate 
the horizontal and vertical 
interactions by writing
\begin{equation}
T_{\mathrm{row}}(L)= T_{v}(L) T_{h}(L) 
\end{equation}
where the matrix elements of 
$T_{v}(L)$ and $T_{h}(L)$ are given by, 
\begin{equation}
\left[T_{\mathrm{v}}(L)\right]_{a_1,\dots,a_L}^{b_1,\dots,b_L}= \prod_{j=1}^{L} W_v(a_j,b_j),~~~ 
\left[T_{\mathrm{h}}(L)\right]_{a_1,\dots,a_L}^{b_1,\dots,b_L}= \prod_{j=1}^{L} W_h(a_j,b_{j+1}) \delta_{a_{j+1},b_{j+1}}
\end{equation}
and for periodic boundary conditions we have $a_{L+1}=a_1$ and $b_{L+1}=b_1$.

As before we can express the partition function of the spin model 
as a trace of a product of transfer matrices,
\begin{equation}
Z_{\mathrm{spin}}(L)= \mathrm{Tr}_{{\cal V}}\left[ T_{\mathrm{row}}(L)\right ]^{L}
\end{equation}

Inspired by our earlier analysis we investigate whether or not 
the transfer matrix of the vertex model
with weights satisfying the condition,
\begin{equation}
\label{mix}
w(i_1,i_2|i_3,i_4)= 0~~\mathrm{for}~~i_1 \neq i_4
\end{equation}
can somehow be related to the spin row-to-row transfer matrix 
by suitable pairing of horizontal and
vertical edge weights. 

As before we have performed this analysis for models up to 
four states per site and with $L \leq 6$. It turns out that we 
find that it is possible 
to match these 
transfer matrices, namely
\begin{equation}
\label{iden}
T_{ver}(L) =T_{row}(L)
\end{equation}
provided that the vertex model weights satisfy the following relation,
\begin{equation}
\label{ansb}
w(i_1,i_2|i_3,i_4)= W_v(i_3,i_1) W_h(i_1,i_2) \delta_{i_1,i_4}
\end{equation}

We remark that specifically for $n=2$ the above result has been 
verified up to $L=10$.
These verifications seems robust enough to conjecture 
the validity of the expressions (\ref{iden},\ref{ansb})
for arbitrary $n$ and $L$. 

\subsection{Application to the Ising model}

We start by presenting the corresponding 
edge weights for the Ising model in a magnetic field. Considering the
spin basis the   
horizontal and vertical edge weights 
are given by,
\begin{eqnarray}
\label{spinwei}
&& W_{h}(+,+)=
e^{\beta \left(J_h+\frac{H}{2}\right)},~~ W_{h}(+,-)=W_{h}(-,+)= e^{-\beta J_h},~~ 
W_{h}(-,-)=
e^{\beta \left(J_h-\frac{H}{2}\right)} \nonumber \\
&& W_{v}(+,+)=
e^{\beta \left(J_v+\frac{H}{2}\right)},~~ W_{v}(+,-)=W_{v}(-,+)= e^{-\beta J_v},~~ 
W_{v}(-,-)=
e^{\beta \left(J_v-\frac{H}{2}\right)}
\end{eqnarray}

We now consider the equivalent vertex model and recall that the respective Boltzmann weights 
satisfy the property
$w(i_1,i_2|i_3,i_4)= 0$ for $i_1 \neq i_4$. This means that we have only eight non-null weights
and by using the graphical representation given in Fig.(\ref{Fig16V}) of the Lax operator (\ref{LAX})
we have,
\begin{eqnarray}
\label{verwei}
&& w_1=w(+,+|+,+),~~w_2=w(-,-|-,-),~~w_5=w(-,+|+,-),~~w_6=w(+,-|-,+) \nonumber \\
&& v_1=w(+,-|+,+),~~v_2=w(-,+|-,-),~~v_5=w(-,-|+,-),~~v_6=w(+,+|-,+) 
\end{eqnarray}

Considering the first equivalence between spin and vertex model (\ref{ansa}) 
it follows from 
Eq.(\ref{spinwei},\ref{verwei}) that the Boltzmann weights of 
the mixed vertex model are,
\begin{eqnarray}
\label{equia}
&& w_1=e^{\beta(J_h+J_v +H)},~~
w_2=e^{\beta(J_h+J_v -H)},~~
w_5=e^{\beta(-J_h+J_v +\frac{H}{2})},~~
w_6=e^{\beta(-J_h+J_v -\frac{H}{2})} \nonumber \\
&& v_1=e^{\beta(J_h-J_v +\frac{H}{2})},~~
v_2=e^{\beta(J_h-J_v -\frac{H}{2})},~~
v_5=v_6=e^{\beta(-J_h-J_v)}
\end{eqnarray}
as have been illustrated in Figure (\ref{FigMap}).

On the other hand the second equivalence between spin 
and vertex model (\ref{ansb}) tell us 
that the corresponding Boltzmann weights of the 
mixed vertex model are,  
\begin{eqnarray}
\label{equib}
&& w_1=e^{\beta(J_h+J_v +H)},~~
w_2=e^{\beta(J_h+J_v -H)},~~
w_5=w_6=e^{\beta(-J_h-J_v)} \nonumber \\
&& v_1=e^{\beta(-J_h+J_v +\frac{H}{2})},~~
v_2=e^{\beta(-J_h+J_v -\frac{H}{2})},~~
v_5=e^{\beta(J_h-J_v -\frac{H}{2})},~~
v_6=e^{\beta(J_h-J_v +\frac{H}{2})}
\end{eqnarray}
which is now illustrated in Fig.(\ref{FigMapextra}).
\setlength{\unitlength}{2500sp}
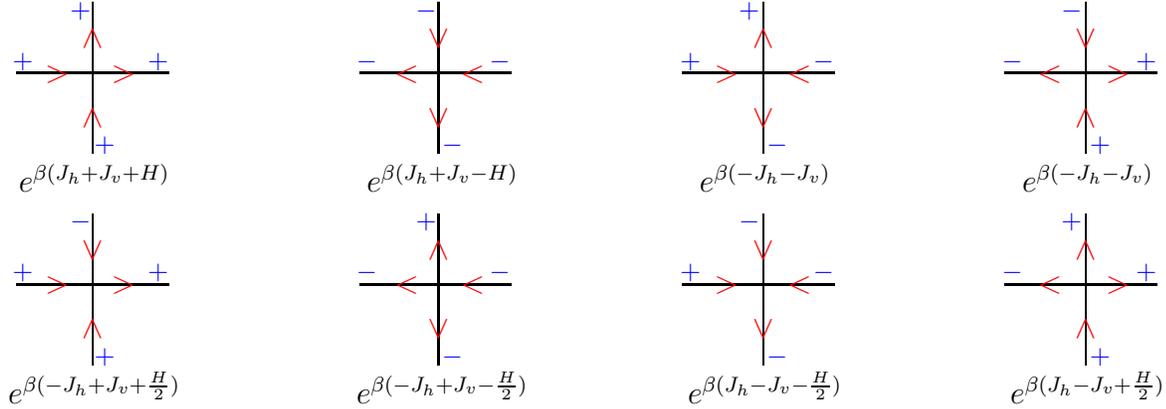
\begin{figure}[ht]
\begin{center}
\begin{picture}(8000,2000)
{\put(-2200,900){\line(1,0){1500}}}
{\put(1200,900){\line(1,0){1500}}}
{\put(4400,900){\line(1,0){1500}}}
{\put(7600,900){\line(1,0){1500}}}
{\put(-2200,-1200){\line(1,0){1500}}}
{\put(1200,-1200){\line(1,0){1500}}}
{\put(4400,-1200){\line(1,0){1500}}}
{\put(7600,-1200){\line(1,0){1500}}}
{\put(-1450,1600){\line(0,-1){1500}}}
{\put(1980,1600){\line(0,-1){1500}}}
{\put(5200,1600){\line(0,-1){1500}}}
{\put(8400,1600){\line(0,-1){1500}}}
{\put(-1450,-500){\line(0,-1){1500}}}
{\put(1980,-500){\line(0,-1){1500}}}
{\put(5200,-500){\line(0,-1){1500}}}
{\put(8400,-500){\line(0,-1){1500}}}
{\put(-1800,890){\makebox(0,0){\fontsize{12}{14}\selectfont $\color{red} >$}}}
{\put(1665,890){\makebox(0,0){\fontsize{12}{14}\selectfont $\color{red} <$}}}
{\put(4835,890){\makebox(0,0){\fontsize{12}{14}\selectfont $\color{red} >$}}}
{\put(8045,890){\makebox(0,0){\fontsize{12}{14}\selectfont $\color{red} <$}}}
{\put(-2140,1010){\makebox(0,0){\fontsize{10}{12}\selectfont $\color{blue} +$}}}
{\put(1270,1010){\makebox(0,0){\fontsize{10}{12}\selectfont $\color{blue} -$}}}
{\put(4480,1010){\makebox(0,0){\fontsize{10}{12}\selectfont $\color{blue} +$}}}
{\put(7670,1010){\makebox(0,0){\fontsize{10}{12}\selectfont $\color{blue} -$}}}
{\put(-1800,-1200){\makebox(0,0){\fontsize{12}{14}\selectfont $\color{red} >$}}}
{\put(1665,-1200){\makebox(0,0){\fontsize{12}{14}\selectfont $\color{red} <$}}}
{\put(4835,-1200){\makebox(0,0){\fontsize{12}{14}\selectfont $\color{red} >$}}}
{\put(8045,-1200){\makebox(0,0){\fontsize{12}{14}\selectfont $\color{red} <$}}}
{\put(-2140,-1080){\makebox(0,0){\fontsize{10}{12}\selectfont $\color{blue} +$}}}
{\put(1270,-1080){\makebox(0,0){\fontsize{10}{12}\selectfont $\color{blue} -$}}}
{\put(4480,-1080){\makebox(0,0){\fontsize{10}{12}\selectfont $\color{blue} +$}}}
{\put(7670,-1080){\makebox(0,0){\fontsize{10}{12}\selectfont $\color{blue} -$}}}
{\put(-1140,890){\makebox(0,0){\fontsize{12}{14}\selectfont $\color{red} >$}}}
{\put(2320,890){\makebox(0,0){\fontsize{12}{14}\selectfont $\color{red} <$}}}
{\put(5560,890){\makebox(0,0){\fontsize{12}{14}\selectfont $\color{red} <$}}}
{\put(8720,890){\makebox(0,0){\fontsize{12}{14}\selectfont $\color{red} >$}}}
{\put(-800,1010){\makebox(0,0){\fontsize{10}{12}\selectfont $\color{blue} +$}}}
{\put(2590,1010){\makebox(0,0){\fontsize{10}{12}\selectfont $\color{blue} -$}}}
{\put(5800,1010){\makebox(0,0){\fontsize{10}{12}\selectfont $\color{blue} -$}}}
{\put(9000,1010){\makebox(0,0){\fontsize{10}{12}\selectfont $\color{blue} +$}}}
{\put(-1140,-1200){\makebox(0,0){\fontsize{12}{14}\selectfont $\color{red} >$}}}
{\put(2320,-1200){\makebox(0,0){\fontsize{12}{14}\selectfont $\color{red} <$}}}
{\put(5560,-1200){\makebox(0,0){\fontsize{12}{14}\selectfont $\color{red} <$}}}
{\put(8720,-1200){\makebox(0,0){\fontsize{12}{14}\selectfont $\color{red} >$}}}
{\put(-800,-1080){\makebox(0,0){\fontsize{10}{12}\selectfont $\color{blue} +$}}}
{\put(2590,-1080){\makebox(0,0){\fontsize{10}{12}\selectfont $\color{blue} -$}}}
{\put(5800,-1080){\makebox(0,0){\fontsize{10}{12}\selectfont $\color{blue} -$}}}
{\put(9000,-1080){\makebox(0,0){\fontsize{10}{12}\selectfont $\color{blue} +$}}}
{\put(-1450,1240){\makebox(0,0){\fontsize{12}{14}\selectfont $\color{red} \wedge$}}}
{\put(1980,1240){\makebox(0,0){\fontsize{12}{14}\selectfont $\color{red} \vee$}}}
{\put(5200,1240){\makebox(0,0){\fontsize{12}{14}\selectfont $\color{red} \wedge$}}}
{\put(8400,1240){\makebox(0,0){\fontsize{12}{14}\selectfont $\color{red} \vee$}}}
{\put(-1570,1520){\makebox(0,0){\fontsize{10}{12}\selectfont $\color{blue} +$}}}
{\put(1860,1520){\makebox(0,0){\fontsize{10}{12}\selectfont $\color{blue} -$}}}
{\put(5060,1520){\makebox(0,0){\fontsize{10}{12}\selectfont $\color{blue} +$}}}
{\put(8260,1520){\makebox(0,0){\fontsize{10}{12}\selectfont $\color{blue} -$}}}
{\put(-1450,-860){\makebox(0,0){\fontsize{12}{14}\selectfont $\color{red} \vee$}}}
{\put(1980,-860){\makebox(0,0){\fontsize{12}{14}\selectfont $\color{red} \wedge$}}}
{\put(5200,-860){\makebox(0,0){\fontsize{12}{14}\selectfont $\color{red} \vee$}}}
{\put(8400,-860){\makebox(0,0){\fontsize{12}{14}\selectfont $\color{red} \wedge$}}}
{\put(-1570,-580){\makebox(0,0){\fontsize{10}{12}\selectfont $\color{blue} -$}}}
{\put(1860,-580){\makebox(0,0){\fontsize{10}{12}\selectfont $\color{blue} +$}}}
{\put(5060,-580){\makebox(0,0){\fontsize{10}{12}\selectfont $\color{blue} -$}}}
{\put(8260,-580){\makebox(0,0){\fontsize{10}{12}\selectfont $\color{blue} +$}}}
{\put(-1450,450){\makebox(0,0){\fontsize{12}{14}\selectfont $\color{red} \wedge$}}}
{\put(1980,450){\makebox(0,0){\fontsize{12}{14}\selectfont $\color{red} \vee$}}}
{\put(5200,450){\makebox(0,0){\fontsize{12}{14}\selectfont $\color{red} \vee$}}}
{\put(8400,450){\makebox(0,0){\fontsize{12}{14}\selectfont $\color{red} \wedge$}}}
{\put(-1330,180){\makebox(0,0){\fontsize{10}{12}\selectfont $\color{blue} +$}}}
{\put(2120,180){\makebox(0,0){\fontsize{10}{12}\selectfont $\color{blue} -$}}}
{\put(5340,180){\makebox(0,0){\fontsize{10}{12}\selectfont $\color{blue} -$}}}
{\put(8540,180){\makebox(0,0){\fontsize{10}{12}\selectfont $\color{blue} +$}}}
{\put(-1450,-1640){\makebox(0,0){\fontsize{12}{14}\selectfont $\color{red} \wedge$}}}
{\put(1980,-1640){\makebox(0,0){\fontsize{12}{14}\selectfont $\color{red} \vee$}}}
{\put(5200,-1640){\makebox(0,0){\fontsize{12}{14}\selectfont $\color{red} \vee$}}}
{\put(8400,-1640){\makebox(0,0){\fontsize{12}{14}\selectfont $\color{red} \wedge$}}}
{\put(-1330,-1920){\makebox(0,0){\fontsize{10}{12}\selectfont $\color{blue} +$}}}
{\put(2120,-1920){\makebox(0,0){\fontsize{10}{12}\selectfont $\color{blue} -$}}}
{\put(5340,-1920){\makebox(0,0){\fontsize{10}{12}\selectfont $\color{blue} -$}}}
{\put(8540,-1920){\makebox(0,0){\fontsize{10}{12}\selectfont $\color{blue} +$}}}
{\put(-1430,-140){\makebox(0,0){\fontsize{12}{14}\selectfont $e^{\beta(J_h+J_v +H)}$}}}
{\put(2020,-140){\makebox(0,0){\fontsize{12}{14}\selectfont $e^{\beta(J_h+J_v -H)}$}}}
{\put(5220,-140){\makebox(0,0){\fontsize{12}{14}\selectfont $e^{\beta(-J_h-J_v)}$}}}
{\put(8420,-140){\makebox(0,0){\fontsize{12}{14}\selectfont $e^{\beta(-J_h-J_v)}$}}}
{\put(-1430,-2220){\makebox(0,0){\fontsize{12}{14}\selectfont $e^{\beta(-J_h+J_v +\frac{H}{2})}$}}}
{\put(2020,-2220){\makebox(0,0){\fontsize{12}{14}\selectfont $e^{\beta(-J_h+J_v -\frac{H}{2})}$}}}
{\put(5220,-2220){\makebox(0,0){\fontsize{12}{14}\selectfont $e^{\beta(J_h-J_v-\frac{H}{2})}$}}}
{\put(8420,-2220){\makebox(0,0){\fontsize{12}{14}\selectfont $e^{\beta(J_h-J_v+\frac{H}{2})}$}}}
\end{picture}
\end{center}
\vspace{2.3cm}
\caption{Alternative equivalence of the Ising model in a magnetic field 
with a mixed 
eight-vertex model on the square lattice
with toroidal boundary conditions.}
\label{FigMapextra}
\end{figure}

\section{Integrable Manifold for the Mixed Vertex Model}

In two spatial dimensions a lattice model of statistical mechanics 
is called integrable 
when the respective transfer matrix commutes 
for distinct set of Boltzmann weights \cite{BAX}. For vertex models   
a sufficient condition for commuting 
transfer matrices 
is the existence of an invertible $\mathrm{R}$-matrix satisfying the following
Yang-Baxter algebra,
\begin{equation}
\label{YBAV}
\mathrm{R}_{12}(\omega^{'},\omega^{''})  
\mathbb{L}_{13}(\omega^{'})
\mathbb{L}_{23}(\omega^{''})=
\mathbb{L}_{23}(\omega^{''})
\mathbb{L}_{13}(\omega{'})
\mathrm{R}_{12}(\omega^{'},\omega^{''}).  
\end{equation}
where $w^{'}$ and $w^{''}$ denote two different 
sets of vertex weights and 
$\mathrm{R}(\omega^{'},\omega^{''})$ denotes 
the $n^2 \times n^2$ $\mathrm{R}$-matrix.   

For the mixed eight-vertex model we may order 
the basis 
as $\ket{+,+}, \ket{+,-}, \ket{-,+}, \ket{-,-}$ 
and the respective Lax operator can be represented 
by the following matrix,
\begin{equation}
\label{LAXn2}
\mathbb{L}^{(\mathrm{mix})}(\omega)=\left[
\begin{array}{cc|cc}
w_1 & 0 & v_1 & 0 \\
v_6 & 0 & w_6 & 0 \\ \hline
0 & w_5 & 0 & v_5 \\
0 & v_2 & 0 & w_2 \\
\end{array}
\right].
\end{equation}

In this section we study solutions 
of the Yang-Baxter equation (\ref{YBAV}) when the 
vertex weights 
are unchanged 
by reversing the arrows which is equivalent to the 
spin reversal
$ + \leftrightarrow -$ symmetry. As a result we have only 
four distinct 
weights due to 
the following identifications,
\begin{equation}
\label{MANI}
w_2=w_1,~~w_6=w_5,~~v_2=v_1,~~v_6=v_5
\end{equation}
and we note  
that such
symmetric manifold encodes 
the Ising model without an external magnetic 
field, see Eqs.(\ref{equia},\ref{equib}).

In what follows we also assume that the underlying $\mathrm{R}$-matrix 
has the same matrix
form of the Lax operator. Indicating its matrix elements
by bold letters we have,
\begin{equation}
\label{RMAn2}
\mathrm{R}^{(\mathrm{mix})}(\omega^{'},\omega^{''})=\left[
\begin{array}{cc|cc}
{\bf w}_1 & 0 & {\bf v}_1 & 0 \\
{\bf v}_6 & 0 & {\bf w}_6 & 0 \\ \hline
0 & {\bf w}_5 & 0 & {\bf v}_5 \\
0 & {\bf v}_2 & 0 & {\bf w}_2 \\
\end{array}
\right],
\end{equation}
where as in the case of the Lax operator we assume ${\bf{w}}_2={\bf{w}}_1$, ${\bf{w}}_6={\bf{w}}_5$,  ${\bf{v}}_2={\bf{v}}_1$ and ${\bf{v}}_6={\bf{v}}_5$.

We obtain the functional equations constraining the $\mathrm{R}$-matrix elements and the 
vertex weights by substituting
the proposals (\ref{LAXn2},\ref{RMAn2}) in the Yang-Baxter equation (\ref{YBAV}). In the case of the symmetric manifold (\ref{MANI})
we have twelve independent relations which can be subdivided in terms of their number of monomials. We have four
simple relations involving only two monomials given by,
\begin{eqnarray}
\label{TWO1}
&& {\bf w}_1 w^{'}_5 w^{''}_1 - {\bf w}_5 w^{'}_1 w^{''}_5=0, \\
\label{TWO2}
&& {\bf w}_5 v^{'}_1 v^{''}_5 - {\bf w}_1 v^{'}_5 v^{''}_1=0, \\
\label{TWO3}
&& {\bf v}_1 w^{'}_5 v^{''}_5 - {\bf v}_5 w^{'}_1 v^{''}_1=0,  \\
\label{TWO4}
&& {\bf v}_5 v^{'}_1 w^{''}_1 - {\bf v}_1 v^{'}_5 w^{''}_5=0, 
\end{eqnarray}

The eight remaining equations contain four monomials and their expressions are,
\begin{eqnarray}
\label{FOUR1}
{\bf w}_5 v^{'}_1 w^{''}_1 - {\bf v}_1 (w^{'}_5 w^{''}_5-v^{'}_1 v^{''}_1) - {\bf w}_1 w^{'}_1 v^{''}_1 =0, \\
\label{FOUR2}
{\bf w}_5 (v^{'}_5 w^{''}_1-w^{'}_1 v^{''}_5) - {\bf v}_5 w^{'}_5 w^{''}_1 + {\bf v}_1 v^{'}_5 v^{''}_1=0, \\
\label{FOUR3}
{\bf w}_1 w^{'}_5 v^{''}_5 - {\bf w}_5 v^{'}_5 w^{''}_5 + {\bf v}_5 (w^{'}_5 w^{''}_5 - v^{'}_1 v^{''}_1)=0, \\
\label{FOUR4}
{\bf w}_1 (v^{'}_5 w^{''}_1 - w^{'}_1 v^{''}_5) + {\bf v}_1 v^{'}_1 v^{''}_5 - {\bf v}_5 w^{'}_1 w^{''}_5=0, \\
\label{FOUR5}
{\bf v}_1 (w^{'}_1 w^{''}_1-v^{'}_5 v^{''}_5) - {\bf w}_1 v^{'}_1 w^{''}_1 + {\bf w}_5 w^{'}_1 v^{''}_1=0, \\
\label{FOUR6}
{\bf v}_1 w^{'}_5 w^{''}_1 - {\bf w}_5 (v^{'}_1 w^{''}_5-w^{'}_5 v^{''}_1) - {\bf v}_5 v^{'}_5 v^{''}_1 =0, \\
\label{FOUR7}
{\bf v}_5 (w^{'}_1 w^{''}_1 - v^{'}_5 v^{''}_5) + {\bf w}_5 w^{'}_5 v^{''}_5 - {\bf w}_1 v^{'}_5 w^{''}_5=0, \\
\label{FOUR8}
{\bf v}_5 v^{'}_1 v^{''}_5 - {\bf v}_1 w^{'}_1 w^{''}_5 + {\bf w}_1 (v^{'}_1 w^{''}_5 - w^{'}_5 v^{''}_1)=0.
\end{eqnarray}
 
We consider the solution of Eqs.(\ref{TWO1}-\ref{FOUR8}) as a system of homogeneous 
polynomials relations for the
$\mathrm{R}$-matrix elements ${\bf w}_1$, ${\bf w}_5$, ${\bf v}_1$ and ${\bf v}_5$. We first 
observe that 
the determinant of
the pair of equations (\ref{TWO1},\ref{TWO2}) and (\ref{TWO3},\ref{TWO4}) are the same and 
such determinant must vanish.
This condition assures that $\mathrm{R}$-matrix entries are not all zero 
and as result we have the constraint,
\begin{equation}
\label{INV1}
\frac{w^{'}_5 v^{'}_1}{w^{'}_1 v^{'}_5}=
\frac{w^{''}_5 v^{''}_1}{w^{''}_1 v^{''}_5}= \Delta_1
\end{equation}
where $\Delta_1$ is a free constant. We now can solve Eqs.(\ref{TWO1},\ref{TWO3}) 
for the $\mathrm{R}$-matrix 
entries ${\bf w}_1 $ and ${\bf v}_1$ to obtain,
\begin{equation}
\label{v0w1v0v1}
{\bf w}_1 = {\bf w}_5 \frac{w^{'}_1 w^{''}_5}{w^{'}_5 w^{''}_1},~~
{\bf v}_1 = {\bf v}_5 \frac{w^{'}_1 v^{''}_1}{w^{'}_5 v^{''}_5}
\end{equation}

By substituting the above results into Eqs.(\ref{FOUR1}-\ref{FOUR4}) we find that they 
provide us a single
functional relation. For instance, we can solve Eqs.(\ref{FOUR1}-\ref{FOUR4}) for ${\bf v}_5 $
and after some simplifications we obtain,
\begin{equation}
\label{v0v5}
{\bf v}_5 = {\bf w}_5 \frac{w^{''}_5 \left(w^{'}_1 v^{''}_5 - v^{'}_5 w^{''}_1\right)}
     {w^{''}_1 \left(v^{'}_1 v^{''}_1 - w^{'}_5 w^{''}_5\right)}
\end{equation}
and consequently the $\mathrm{R}$-matrix entries have the same common factor ${\bf w}_5$.

At this point we are left to solve Eqs.(\ref{FOUR5}-\ref{FOUR8}). After using the
previous results one can show that these relations are proportional to each other.
We find that their solution gives rise to a second algebraic invariant,
\begin{equation}
\label{INV2}
\frac{(w^{'}_1)^2+(v^{'}_5)^2-(w^{'}_5)^2-(v^{'}_1)^2}{2 w^{'}_1 v^{'}_5}=
\frac{(w^{''}_1)^2+(v^{''}_5)^2-(w^{''}_5)^2-(v^{''}_1)^2}{2 w^{''}_1 v^{''}_5}= \Delta_2
\end{equation}
where $\Delta_2$ is a free constant.

Interesting enough, the algebraic invariants 
(\ref{INV1},\ref{INV2}) 
have a direct one-to-one correspondence
with those associated to 
the even eight-vertex model 
with symmetric weights. This equivalence
is summarized in Appendix A and as result one concludes
that the vertex weights $w_1$, $w_5$, $v_1$, $v_5$ sit 
on the same elliptic curve satisfied by the vertex weights
of the even eight-vertex model \cite{BAX}. Following
Baxter monograph \cite{BAX} the vertex weights 
of the symmetric mixed
eight-vertex model can be uniformized in terms of Jacobi
elliptic functions,
\begin{eqnarray}
\label{UNIF}
&& w_1(x)= sn[x+i\lambda,k],~~w_5(x)=sn[i\lambda,k], \nonumber \\
&& v_5(x)=sn[x,k],~~v_1(x)=-k sn[i\lambda,k] sn[x,k] sn[x+i\lambda,k]
\end{eqnarray}
where $x$ is the spectral parameter, $\lambda$ is a free parameter 
and $sn[x,k]$ represents 
the elliptic Jacobi function of modulus $k$. The dependence 
of the invariants $\Delta_1$ and $\Delta_2$ with the 
modulus and $\lambda$ is
given by,
\begin{equation}
\Delta_1=-k sn[i\lambda,k]^2,~~\Delta_2=cn[i\lambda,k] dn[i\lambda,k]
\end{equation}
where $cn[x,k]$ and $dn[x,k]$ denote the other two 
elliptic Jacobi functions. 

However, contrary to what happens with the symmetric 
even eight-vertex model, the $\mathrm{R}$-matrix associated 
to the
mixed symmetric eight-vertex model is not given 
as a function
of the difference of spectral parameters. We can see that
computing explicitly the
$\mathrm{R}$-matrix elements 
(\ref{v0w1v0v1},\ref{v0v5}) in terms of the
spectral parameters $x_1$ and $x_2$ associated 
to the uniformization 
of the set of weights 
\{ $w^{'}_1$, $w^{'}_5$, $v^{'}_1$, $v^{'}_5$ \} and 
\{ $w^{''}_1$, $w^{''}_5$, $v^{''}_1$, $v^{''}_5$ \}, respectively. 
Taking into account the uniformization (\ref{UNIF}) and 
with the
help of addition identities of elliptic functions we obtain,
\begin{equation}
\frac{\bf w_1}{\bf w_5}= \frac{sn(i\lambda+x_1,k)}{sn(i\lambda+x_2,k)},~~
\frac{\bf v_5}{\bf w_5}= \frac{sn(x_1-x_2,k)}{sn(i\lambda+x_2,k)},~~
\frac{\bf v_1}{\bf w_5}= -k sn(i\lambda+x_1,k) sn(x_1-x_2,k)
\end{equation}
and 
we observe that not all the matrix elements can be represented solely  as 
functions 
of the difference of
spectral parameters. This means that the $\mathrm{R}$-matrix of the mixed symmetric
eight-vertex model lie on a surface rather than a curve such is the
case of the $\mathrm{R}$-matrix of the symmetric even eight-vertex model. See
appendix A for a discussion about the geometry of 
the later $\mathrm{R}$-matrix. 

If we choose the overall normalization 
${\bf w}_5=\frac{1}{1+i\sqrt{k}sn(x_1-x_2,k)}$ 
one can show 
that the $\mathrm{R}$-matrix of the mixed 
eight-vertex model 
with symmetric weights
satisfies the standard 
unitarity property,
\begin{equation}
\label{UNI}
\mathrm{R}^{\mathrm{(mix)}}_{12}(x_1,x_2) \mathrm{R}^{\mathrm{(mix)}}_{21}(x_2,x_1)= \mathrm{I_4}
\end{equation}
reducing to the $4 \times 4$ permutator at the point $x_2=x_1$.

In addition to that, we have also verified 
that such $\mathrm{R}$-matrix 
fulfills the Yang-Baxter equation,
\begin{equation}
\label{YANBAX}
\mathrm{R}^\mathrm{(mix)}_{12}(x_1,x_2) 
\mathrm{R}^\mathrm{(mix)}_{13}(x_1,x_3) 
\mathrm{R}^\mathrm{(mix)}_{23}(x_2,x_3)=
\mathrm{R}^\mathrm{(mix)}_{23}(x_2,x_3) 
\mathrm{R}^\mathrm{(mix)}_{13}(x_1,x_3)
\mathrm{R}^\mathrm{(mix)}_{12}(x_1,x_2) 
\end{equation}
being a sufficient condition 
for the associativity 
of the Yang-Baxter algebra.

Our last remark concerns with the commutativity of the 
transfer matrix of the 
Ising model with zero magnetic
field. This spin model is encoded in the symmetric 
mixed eight-vertex 
and considering the first spin-vertex 
equivalence with $H=0$
we have, 
\begin{equation}
\label{equiaZERO}
w_1=w_2=e^{\beta(J_h+J_v)},~~
w_5=w_6=e^{\beta(-J_h+J_v)},~~
v_1=v_2=e^{\beta(J_h-J_v) },~~
v_5=v_6=e^{\beta(-J_h-J_v)}
\end{equation}

At this point we recall the transfer matrix of the mixed eight-vertex model 
commutes when the weights
satisfy the restrictions (\ref{INV1},\ref{INV2}). 
By substituting the weights (\ref{equiaZERO}) 
in the relation (\ref{INV1}) 
we find that the invariant $\Delta_1$ has is fixed to the unity. 
The second invariant $\Delta_2$ provides us a relation among the
horizontal 
and vertical 
spin couplings given by,
\begin{equation}
\Delta_2=2 \sinh(2\beta J_h) \sinh(2\beta J_v)
\end{equation}
reproducing the celebrated condition for the 
commutativity of 
the diagonal-to-diagonal 
transfer matrix of the Ising model 
with zero magnetic field \cite{ONSA}. 
Recall here that 
such relation has been previously
derived in the context of the 
even eight-vertex which
can be regarded as a two next 
nearest-neighbour Ising model at some 
decoupling point \cite{BAX}.
By way of contrast our 
derivation is in the context of the 
Ising model 
originally solved by Onsager \cite{ONSA1}. 

It turns out that similar analysis can also 
be carried for the 
second spin-vertex 
correspondence (\ref{equib}) which 
for $H=0$ reads,
\begin{equation}
\label{equibZERO}
w_1=w_2=e^{\beta(J_h+J_v)},~~
w_5=w_6=e^{\beta(-J_h-J_v)},~~
v_1=v_2=e^{\beta(-J_h+J_v)},~~
v_5=v_6=e^{\beta(J_h-J_v)}
\end{equation}

Now after substituting the weights (\ref{equibZERO}) 
in the expressions
of  the invariants (\ref{INV1},\ref{INV2}) we obtain,
\begin{equation}
\Delta_1=e^{4 \beta J_h},~~ \Delta_2= 2 e^{-2\beta J_h} \cosh(2\beta J_v) \sinh(2 \beta J_h)
\end{equation}
and therefore the row-to-row transfer matrix of the Ising model 
does not yields 
a one parameter family of commuting operators.

\subsection{The $\mathrm{R}$-matrix geometry}

In order to investigate the geometric properties of the 
$\mathrm{R}$-matrix 
of the symmetric mixed eight-vertex
model we need to determine  
the form algebraic variety  which is satisfied by 
matrix elements 
${\bf w_1}$, ${\bf w_5}$, ${\bf v_1}$ and ${\bf v_5}$. To this end we first analyze 
the behaviour
of the left hand side 
of algebraic invariants (\ref{INV1},\ref{INV2}) when the vertex weights are replaced
by the respective $\mathrm{R}$-matrix elements. More precisely, want to compute the
auxiliary functions $F_1(\omega^{'},\omega^{''})$ and $F_2(\omega^{'},\omega^{''})$ such that,
\begin{equation}
\label{INVdois}
\frac{{\bf w}_5 {\bf v}_1}{{\bf w}_1 {\bf v}_5}=F_1(\omega^{'},\omega^{''}),~~ 
\frac{({\bf w}_1)^2+({\bf v}_5)^2-(\bf{w_5})^2-({\bf v_1})^2}{2 {\bf w_1} {\bf v_5}}=F_2(\omega^{'},\omega^{''})
\end{equation}

We substitute in Eq.(\ref{INVdois}) the expressions of 
the $\mathrm{R}$-matrix 
elements (\ref{v0w1v0v1},\ref{v0v5}) and after a systematic use 
of the vertex
weights invariants (\ref{INV1},\ref{INV2}) we find,
\begin{equation}
\label{FUNdois}
F_1(\omega^{'},\omega^{''})= \frac{ w^{''}_1 v^{''}_1}{ w^{''}_5 v^{''}_5},~~
F_2(\omega^{'},\omega^{''})= 
\frac{(w^{''}_1)^2+(v^{''}_1)^2-(w^{''}_5)^2-(v^{''}_5)^2}{2 w^{''}_5 v^{''}_5}
\end{equation}

The fact that functions 
$F_1(\omega^{'},\omega^{''})$ and $F_2(\omega^{'},\omega^{''})$ 
are not constants 
but instead vertex weights dependent suggests that the variables 
${\bf w_1}$, ${\bf w_5}$, ${\bf v_1}$, ${\bf v_5}$ should lie
on two-dimensional variety. The algebraic form  of such surface can 
be determined 
after we eliminate the
vertex weights $w^{''}_1,w^{''}_2,v^{''}_1,v^{''}_5$ of Eqs.(\ref{INVdois},\ref{FUNdois}). 
The technical details of this computation are summarized in Appendix B and in what
follows we present the main result. It turns out that the underlying surface 
is defined
by the following homogeneous quartic polynomial,
\begin{eqnarray}
\label{SURF}
S &=& ({\bf v}_1)^4 + 4\frac{\Big(1 + (\Delta_1)^2 - (\Delta_2)^2\Big)}{\Delta_1} {\bf v}_1 {\bf v}_5 {\bf w}_1 {\bf w}_5  
-2({\bf v}_1)^2 \Big(({\bf v}_5)^2 + ({\bf w}_1)^2 + ({\bf w}_5)^2\Big) \nonumber \\
&+&  ({\bf v}_5 - {\bf w}_1 - {\bf w}_5) ({\bf v}_5 + {\bf w}_1 - {\bf w}_5) ({\bf v}_5 - {\bf w}_1 + {\bf w}_5)
     ({\bf v}_5 + {\bf w}_1 + {\bf w}_5) 
\end{eqnarray}

The geometric properties of surfaces has been studied by a number of algebraic 
geometers long ago and have culminated in
the famous Kodaira-Enriques classification, see for instance \cite{BIU,HUL}. The crucial
point in such classification problem concerns with the resolution of the nature 
of the surface singularities. 
The singular points on a surface
form a closed subvariety $\mathrm{Sing}(S)$ determined by
the zeroes of all the partial 
derivatives of $S$, namely
\begin{equation}
\label{sing}
\mathrm{Sing}(S)=\left\{[{\bf w}_1:{\bf w}_5:{\bf v}_1: {\bf v}_5]
\in \mathbb{CP}^3 {\Big | }\frac{\partial S}{\partial {\bf w}_1}=0,\frac{\partial S}{\partial {\bf w}_5}=0,\frac{\partial S}{\partial {\bf v}_1}=0,\frac{\partial S}{\partial {\bf v}_5}=0
\right\} 
\end{equation}
and we find that $\mathrm{Sing}(S)$ is constituted of twelve isolated singular points given by,
\begin{eqnarray}
&& P_1^{\pm}=[0:1:0:\pm 1],~~ P_2^{\pm}=[0:1:\pm 1:0],~~ P_3^{\pm}=[1:\pm 1: 0:0] \nonumber \\
&& P_4^{\pm}=[1:0:\pm 1:0],~~ P_5^{\pm}=[1:0:0:\pm 1],~~ P_6^{\pm}=[0:0: 1:\pm 1] 
\end{eqnarray}

Here we are in a fortunate situation since the presence of only 
isolated singularities 
tell us that $S$ is a normal surface. It turns out that a normal 
quartic 
surface can be 
either a $\mathrm{K3}$ surface, a ruled surface over 
an elliptic or a genus 3 curves
or still a rational surface \cite{UME,URA}. The classification 
problem 
of the surface (\ref{SURF})
in one of these
four possible categories can be done investigating the nature of 
the underlying singularities. In our case we find that all the 
twelve singularities
are ordinary double points since the Taylor series 
expansion around
the singular points give rise to nondegenerate 
quadratic forms \cite{ANAR}. These type of singularities 
do not affect 
the geometric properties
of the surface and the minimal resolution of the singularities 
are non-singular 
quartics \cite{UME,URA}. It is well known that projective quartic 
surfaces without singularities 
are classical examples of 
the $\mathrm{K3}$ surfaces \cite{BIU,HUL} and therefore 
the birational class of the surface $S$ is characterized
as follows,
\begin{equation}
S \setminus \mathrm{Sing}(S) \cong \mathrm{K3}~~\mathrm{surface}
\end{equation}

We finally stress that this scenario is very different 
from that of the even 
eight-vertex model with symmetry weights. In fact, the 
$\mathrm{R}$-matrix of the symmetric even eight-vertex model
lie on a elliptic curve rather on a surface, see Appendix A.

\subsection{The Hamiltonian Limit}

Here we consider the Hamiltonian limit of the integrable mixed 
eight-vertex with symmetry weights discussed in the previous section. The two-body Hamiltonian $H^{(\mathrm{mix})}_{j,j+1}$
is obtained by expanding the 
respective Lax operator around the 
permutation operator $P$ and up to the first order we have,
\begin{equation}
L^{(\mathrm{mix})}(\omega)=P \left( 1+ \varepsilon H^{(\mathrm{mix})}_{j,j+1} \right),~~~
\end{equation}
where $\varepsilon$ is the expansion parameter 
and the permutator $P =\displaystyle \sum_{i_1,i_2=1}^{2} e_{i_1,i_2}^{(j)} \otimes e_{i_2,i_1}^{(j+1)}$.

The expansion of the 
vertex weights reducing the Lax operator to the permutator 
at zero order 
is as follows,
\begin{equation}
{w}_1=1+\varepsilon {\dot{w}}_1,~~
{w}_5=1+\varepsilon {\dot{w}}_5,~~
{v}_1=\varepsilon {\dot{v}}_1,~~
{v}_5=\varepsilon {\dot{v}}_5
\end{equation}
and since 
the vertex weights are require to satisfy 
the invariants (\ref{INV1},\ref{INV2}) we have two
constraints among the expansion
coefficients,
\begin{equation}
{\dot{v}}_1 =\Delta_1 {\dot{v}}_5,~~
{\dot{w}}_1 -{\dot{w}}_5=\Delta_2 {\dot{v}}_5
\end{equation}

Collecting the above results we find that the
two-body Hamiltonian can be represented by the 
following matrix,
\begin{equation}
\label{HTWOMIX}
{H}^{(\mathrm{mix})}_{j,j+1}=\left[
\begin{array}{cc|cc}
{\dot w}_5 + \Delta_2{\dot v}_5& 0 & \Delta_1{\dot v}_5 & 0 \\
0 &  {\dot w}_5 & 0& {\dot v}_5 \\ \hline
{\dot v}_5 & 0 & {\dot w}_5 & 0 \\
0 & \Delta_1 {\dot v}_5 & 0 &  {\dot w}_5 + \Delta_2{\dot v}_5 \\
\end{array}
\right]_{j,j+1}
\end{equation}

The resulting Hamiltonian for a chain of length $L$ can 
be represented 
in terms of
spin-$\frac{1}{2}$ Pauli matrices, 
\begin{equation}
\sigma^{x}= \left( \begin{array}{cc} 0 & 1 \\ 1 & 0 \\ \end{array} \right),~~
\sigma^{y}= \left( \begin{array}{cc} 0 & -i \\ i & 0 \\ \end{array} \right),~~
\sigma^{z}= \left( \begin{array}{cc} 1 & 0 \\ 0 & -1 \\ \end{array} \right),~~
\end{equation}
and by choosing the overall normalization
${\dot v}_5=-2J/\Delta_2$ we find that the expression 
of the Hamiltonian 
up to an additive constant is, 
\begin{equation}
\label{HAMMIXTOT}
H^{(\mathrm{mix})}= -J\sum_{j=1}^{L} \left(\sigma^{z}_{j} \sigma^{z}_{j+1} +\left( \frac{\Delta_1+1}{\Delta_2} \right) \sigma^{x}_j 
+i\left(\frac{\Delta_1-1}{\Delta_2} \right) \sigma^y_{j}  \sigma^{z}_{j+1} \right)
\end{equation}
where periodic boundary condition $\sigma^{(x,y,z)}_{L+1} \equiv \sigma^{(x,y,z)}_1$ 
is assumed. 

Since the Pauli matrices are Hermitian the Hamiltonian (\ref{HAMMIXTOT}) 
is non-Hermitian 
for arbitrary
values of the parameters $\Delta_1$ and $\Delta_2$. However, this operator 
becomes Hermitian 
if we restrict 
the parameters to the following subspace,
\begin{equation}
\Delta_1=\exp(-i2\theta),~~ \Delta_2=\frac{2\exp(-i\theta)}{\kappa}
\end{equation}
with $0 \leq \theta \leq \pi$ and $\kappa \in \mathbb{R}$.
In this situation we can rewrite the Hamiltonian as, 
\begin{equation}
\label{HAMMIXTOT1}
H^{(\mathrm{mix})}= -J\sum_{j=1}^{L} \left( \sigma^{z}_{j} \sigma^{z}_{j+1} +h\sigma^{x}_j +D\sigma^y_{j} \sigma^{z}_{j+1} \right)
\end{equation}
where the couplings $h$ and $D$ lie on a circle of radius $\kappa$,
\begin{equation}
h=\kappa \cos(\theta),~~ D=\kappa \sin(\theta)
\end{equation}

The first two terms of the Hamiltonian (\ref{HAMMIXTOT1}) 
represent 
the Ising quantum spin chain in a transverse field interaction
related to the classical two-dimensional Ising model 
in the absence 
of a magnetic field. The third term resembles 
the type of exchange interaction devised by 
Dzyaloshinsky and Moriya \cite{DZMO}  to explain the
phenomenon of weak ferromagnetism. This can be better seen
by considering the following canonical transformation 
on the Pauli matrices,
\begin{equation}
\label{change1}
\sigma_{j}^{x} \rightarrow \sigma_{j}^{z},~~ 
\sigma_{j}^{y} \rightarrow {\mathrm U}\sigma_{j}^{y} +{\mathrm V}\sigma_{j}^{x},~~
\sigma_{j}^{z} \rightarrow {\mathrm V}\sigma_{j}^{y} -{\mathrm U}\sigma_{j}^{x},~~
\end{equation}
where the transformation parameters satisfy the relations,
\begin{equation}
\label{change2}
{\mathrm U}^2+{\mathrm V}^2=1,~~ 
{\mathrm U}=\frac{\sqrt{1+iD}+\sqrt{1-iD}}{2(1+D^2)^{1/4}}
\end{equation}

By performing the transformation defined by 
Eqs.(\ref{change1},\ref{change2}) 
we find that the 
Hamiltonian (\ref{HAMMIXTOT1}) can be
rewritten as follows,
\begin{equation}
\label{HAMMIXTOT2}
H^{(\mathrm{mix})}= -J\sum_{j=1}^{L} \left( \frac{(1+\gamma)}{2} \sigma^{x}_{j} \sigma^{x}_{j+1} 
+\frac{(1-\gamma)}{2} \sigma^{y}_{j} \sigma^{y}_{j+1} 
+h\sigma^{z}_j +\frac{D}{2}(\sigma^x_{j} \sigma^{y}_{j+1} -\sigma^y_{j} \sigma^{x}_{j+1}) \right)
\end{equation}
where the coupling $\gamma=\sqrt{1+D^2}$. 

We note that the first two terms 
of Eq.(\ref{HAMMIXTOT2}) 
represent the 
Hamiltonian of the
$XY$ model in the presence of a perpendicular magnetic
field $h$. On the other hand,  
the last antisymmetric 
term is exactly the interaction we obtain when
the Dzyaloshinsky-Moriya exchange vector is 
projected 
in the $z$-direction.

We conclude this section with the following remark. 
It is well known that 
the classical statistical model having the $XY$ model with a magnetic 
field in the $z$-direction
as the underlying spin chain turns out to be  
the even eight-vertex 
model with weights satisfying the free-fermion 
condition \cite{KR,FEL,KAS}. The Hamiltonian limit of this vertex model
has an arbitrary choice of one free parameter in the weights expansion
which gives rise to 
the Dzyaloshinsky-Moriya interaction. In fact, for periodic boundary conditions
the Dzyaloshinsky-Moriya term commutes with the Hamiltonian of the $XY$ spin chain.
Since the Hamiltonian
limits of the mixed eight-vertex and even free-fermion 
eight-vertex 
models are
somehow related it is natural to ask whether or not 
such relationship
can be extended on the level 
of the corresponding partition functions.
In next section we explore this possibility and argued 
that this is
case provided that the weights of the mixed eight-vertex
model satisfy certain constraints.

\section{Mapping among eight-vertex models}

The partition function of vertex models may be 
invariant under
various kind of transformations between the 
corresponding Boltzmann weights. One important such 
symmetry is called
gauge transformation which a similarity 
transformation acting on the horizontal and vertical space
of states of the vertex model \cite{WEG}. 
Here we shall study such transformation for the mixed 
eight-vertex
model which in terms of the respective  Lax operator reads,
\begin{equation}
\label{gauge}
\mathbb{L}^{(\mathrm{tra})}(\omega)= \left (M_1 \otimes M_2 \right)  \left[
\begin{array}{cc|cc}
w_1 & 0 & v_1 & 0 \\
v_6 & 0 & w_6 & 0 \\ \hline
0 & w_5 & 0 & v_5 \\
0 & v_2 & 0 & w_2 \\
\end{array}
\right] \left(M_1 \otimes M_2 \right)^{-1}
\end{equation}
where $M_1$ and $M_2$ may be any non-singular 
$2 \times 2$ matrices.

We consider that the transformed Lax operator 
$\mathbb{L}^{(\mathrm{tra})}(\omega)$ 
to be in the 
form of 
that associated
to the even eight-vertex. In order to avoid 
confusion 
between
weights notations we write the Lax operator 
of the even 
eight-vertex model as,
\begin{equation}
\label{8VEReven}
\mathbb{L}^{(\mathrm{tra})}(\omega) \equiv 
\mathbb{L}^{(\mathrm{even})}(\omega)=
\left[ \begin{array}{cc|cc}
a_+ & 0 & 0 & d_+ \\
0 & b_+ & c_+ & 0 \\ \hline
0 & c_- & b_- & 0 \\
d_- & 0 & 0 & a_- \\
\end{array} \right]
\end{equation}
where for completeness the vertex configurations are illustrated in Fig.(\ref{8even}).
\setlength{\unitlength}{2500sp}
\begin{figure}[ht]
\begin{center}
\begin{picture}(8000,2000)
{\put(-2200,900){\line(1,0){1400}}}
{\put(-600,900){\line(1,0){1400}}}
{\put(1000,900){\line(1,0){1400}}}
{\put(2600,900){\line(1,0){1400}}}
{\put(4200,900){\line(1,0){1400}}}
{\put(5800,900){\line(1,0){1400}}}
{\put(7400,900){\line(1,0){1400}}}
{\put(9000,900){\line(1,0){1400}}}
{\put(-1450,1600){\line(0,-1){1400}}}
{\put(138,1600){\line(0,-1){1400}}}
{\put(1740,1600){\line(0,-1){1400}}}
{\put(3338,1600){\line(0,-1){1400}}}
{\put(4940,1600){\line(0,-1){1400}}}
{\put(6540,1600){\line(0,-1){1400}}}
{\put(8140,1600){\line(0,-1){1400}}}
{\put(9740,1600){\line(0,-1){1400}}}
{\put(-1800,890){\makebox(0,0){\fontsize{12}{14}\selectfont $\color{red} >$}}}
{\put(-215,890){\makebox(0,0){\fontsize{12}{14}\selectfont $\color{red} <$}}}
{\put(1385,890){\makebox(0,0){\fontsize{12}{14}\selectfont $\color{red} >$}}}
{\put(2985,890){\makebox(0,0){\fontsize{12}{14}\selectfont $\color {red} <$}}}
{\put(4585,890){\makebox(0,0){\fontsize{12}{14}\selectfont $\color{red} >$}}}
{\put(6185,890){\makebox(0,0){\fontsize{12}{14}\selectfont $\color{red} <$}}}
{\put(7785,890){\makebox(0,0){\fontsize{12}{14}\selectfont $\color{red} <$}}}
{\put(9350,890){\makebox(0,0){\fontsize{12}{14}\selectfont $\color{red} >$}}}
{\put(-2140,1010){\makebox(0,0){\fontsize{10}{12}\selectfont $\color{blue} +$}}}
{\put(-530,1010){\makebox(0,0){\fontsize{10}{12}\selectfont $\color{blue} -$}}}
{\put(1070,1010){\makebox(0,0){\fontsize{10}{12}\selectfont $\color{blue} +$}}}
{\put(2650,1010){\makebox(0,0){\fontsize{10}{12}\selectfont $\color{blue} -$}}}
{\put(4250,1010){\makebox(0,0){\fontsize{10}{12}\selectfont $\color{blue} +$}}}
{\put(5850,1010){\makebox(0,0){\fontsize{10}{12}\selectfont $\color{blue} -$}}}
{\put(7450,1010){\makebox(0,0){\fontsize{10}{12}\selectfont $\color{blue} -$}}}
{\put(9015,1010){\makebox(0,0){\fontsize{10}{12}\selectfont $\color{blue} +$}}}
{\put(-1140,890){\makebox(0,0){\fontsize{12}{14}\selectfont $\color{red} >$}}}
{\put(460,890){\makebox(0,0){\fontsize{12}{14}\selectfont $\color{red} <$}}}
{\put(2060,890){\makebox(0,0){\fontsize{12}{14}\selectfont $\color{red} >$}}}
{\put(3660,890){\makebox(0,0){\fontsize{12}{14}\selectfont $\color{red} <$}}}
{\put(5260,890){\makebox(0,0){\fontsize{12}{14}\selectfont $\color{red} <$}}}
{\put(6860,890){\makebox(0,0){\fontsize{12}{14}\selectfont $\color{red} >$}}}
{\put(8460,890){\makebox(0,0){\fontsize{12}{14}\selectfont $\color{red} >$}}}
{\put(10060,890){\makebox(0,0){\fontsize{12}{14}\selectfont $\color{red} <$}}}
{\put(-920,1010){\makebox(0,0){\fontsize{10}{12}\selectfont $\color{blue} +$}}}
{\put(700,1010){\makebox(0,0){\fontsize{10}{12}\selectfont $\color{blue} -$}}}
{\put(2300,1010){\makebox(0,0){\fontsize{10}{12}\selectfont $\color{blue} +$}}}
{\put(3900,1010){\makebox(0,0){\fontsize{10}{12}\selectfont $\color{blue} -$}}}
{\put(5500,1010){\makebox(0,0){\fontsize{10}{12}\selectfont $\color{blue} -$}}}
{\put(7100,1010){\makebox(0,0){\fontsize{10}{12}\selectfont $\color{blue} +$}}}
{\put(8700,1010){\makebox(0,0){\fontsize{10}{12}\selectfont $\color{blue} +$}}}
{\put(10300,1010){\makebox(0,0){\fontsize{10}{12}\selectfont $\color{blue} -$}}}
{\put(-1450,1240){\makebox(0,0){\fontsize{12}{14}\selectfont $\color{red} \wedge$}}}
{\put(140,1240){\makebox(0,0){\fontsize{12}{14}\selectfont $\color{red} \vee$}}}
{\put(1740,1240){\makebox(0,0){\fontsize{12}{14}\selectfont $\color{red} \vee$}}}
{\put(3340,1240){\makebox(0,0){\fontsize{12}{14}\selectfont $\color{red} \wedge$}}}
{\put(4940,1240){\makebox(0,0){\fontsize{12}{14}\selectfont $\color{red} \wedge$}}}
{\put(6540,1240){\makebox(0,0){\fontsize{12}{14}\selectfont $\color{red} \vee$}}}
{\put(8150,1240){\makebox(0,0){\fontsize{12}{14}\selectfont $\color{red} \wedge$}}}
{\put(9750,1240){\makebox(0,0){\fontsize{12}{14}\selectfont $\color{red} \vee$}}}
{\put(-1570,1520){\makebox(0,0){\fontsize{10}{12}\selectfont $\color{blue} +$}}}
{\put(20,1520){\makebox(0,0){\fontsize{10}{12}\selectfont $\color{blue} -$}}}
{\put(1620,1520){\makebox(0,0){\fontsize{10}{12}\selectfont $\color{blue} -$}}}
{\put(3220,1520){\makebox(0,0){\fontsize{10}{12}\selectfont $\color{blue} +$}}}
{\put(4820,1520){\makebox(0,0){\fontsize{10}{12}\selectfont $\color{blue} +$}}}
{\put(6420,1520){\makebox(0,0){\fontsize{10}{12}\selectfont $\color{blue} -$}}}
{\put(8030,1520){\makebox(0,0){\fontsize{10}{12}\selectfont $\color{blue} +$}}}
{\put(9630,1520){\makebox(0,0){\fontsize{10}{12}\selectfont $\color{blue} -$}}}
{\put(-1450,540){\makebox(0,0){\fontsize{12}{14}\selectfont $\color{red} \wedge$}}}
{\put(140,540){\makebox(0,0){\fontsize{12}{14}\selectfont $\color{red} \vee$}}}
{\put(1740,540){\makebox(0,0){\fontsize{12}{14}\selectfont $\color{red} \vee$}}}
{\put(3340,540){\makebox(0,0){\fontsize{12}{14}\selectfont $\color{red} \wedge$}}}
{\put(4940,540){\makebox(0,0){\fontsize{12}{14}\selectfont $\color{red} \vee$}}}
{\put(6540,540){\makebox(0,0){\fontsize{12}{14}\selectfont $\color{red} \wedge$}}}
{\put(8150,540){\makebox(0,0){\fontsize{12}{14}\selectfont $\color{red} \vee$}}}
{\put(9750,540){\makebox(0,0){\fontsize{12}{14}\selectfont $\color{red} \wedge$}}}
{\put(-1330,280){\makebox(0,0){\fontsize{10}{12}\selectfont $\color{blue} +$}}}
{\put(260,280){\makebox(0,0){\fontsize{10}{12}\selectfont $\color{blue} -$}}}
{\put(1860,280){\makebox(0,0){\fontsize{10}{12}\selectfont $\color{blue} -$}}}
{\put(3450,280){\makebox(0,0){\fontsize{10}{12}\selectfont $\color{blue} +$}}}
{\put(5050,280){\makebox(0,0){\fontsize{10}{12}\selectfont $\color{blue} -$}}}
{\put(6650,280){\makebox(0,0){\fontsize{10}{12}\selectfont $\color{blue} +$}}}
{\put(8270,280){\makebox(0,0){\fontsize{10}{12}\selectfont $\color{blue} -$}}}
{\put(9870,280){\makebox(0,0){\fontsize{10}{12}\selectfont $\color{blue} +$}}}
{\put(-1430,-50){\makebox(0,0){\fontsize{12}{14}\selectfont $a_+$}}}
{\put(160,-50){\makebox(0,0){\fontsize{12}{14}\selectfont $a_-$}}}
{\put(1770,-50){\makebox(0,0){\fontsize{12}{14}\selectfont $b_-$}}}
{\put(3370,-50){\makebox(0,0){\fontsize{12}{14}\selectfont $b_+$}}}
{\put(4970,-50){\makebox(0,0){\fontsize{12}{14}\selectfont $c_-$}}}
{\put(6580,-50){\makebox(0,0){\fontsize{12}{14}\selectfont $c_+$}}}
{\put(8190,-50){\makebox(0,0){\fontsize{12}{14}\selectfont $d_-$}}}
{\put(9780,-50){\makebox(0,0){\fontsize{12}{14}\selectfont $d_+$}}}
\end{picture}
\end{center}
\caption{The configurations of the even eight-vertex model with weights $a_{\pm},b_{\pm},c_{\pm}$ and $d_{\pm}$.}
\label{8even}
\end{figure}
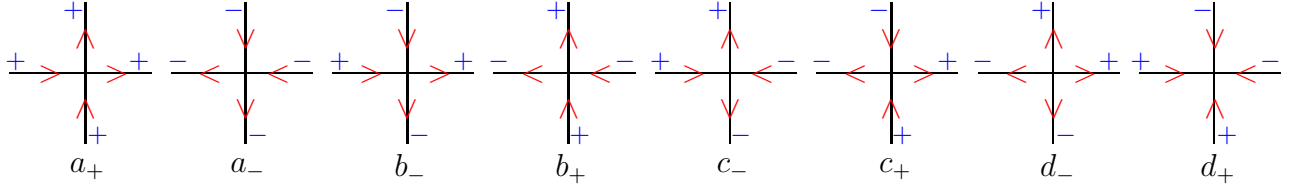

The gauge transformation (\ref{gauge}) leads to eight quadratic polynomial 
relations among 
the elements of the matrices $M_1$ and $M_2$ which have to be solved before 
the weights $a_{\pm},b_{\pm},c_{\pm}, d_{\pm}$ are fixed. These relations are 
directly associated to the fact that the even eight-vertex models have 
exactly eight null weights.
We find that these equations
have a solution provided that the weights of the mixed eight-vertex model 
satisfy the 
following constraints,
\begin{equation}
\label{constra}
w_2=w_1, ~~~v_1 v_6 w_5= v_2 v_5 w_6
\end{equation}
and the corresponding transformation matrices are given by,
\begin{equation}
M_1=
\left( \begin{array}{cc}
1 & \sqrt{\frac{v1}{v2}} \\
-z_1\sqrt{\frac{v2}{v1}} & z_1\\
\end{array} \right),~~
M_2=
\left( \begin{array}{cc}
1 & \sqrt{\frac{v5}{v6}} \\
-z_2\sqrt{\frac{v6}{v5}} & z_2\\
\end{array} \right),~~
\end{equation}
where $z_1$ and $z_2$ are arbitrary non-null free parameters.

The remaining eight relations coming 
from the gauge transformation 
are able to
determine the weights of the equivalent 
even eight-vertex model. After some simplifications we obtain,
\begin{eqnarray}
&& a_{\pm}=\frac{w_1}{2}+\frac{\sqrt{w_6 w_5}}{2} \pm \frac{\sqrt{v_1 v_2}}{2} \pm \frac{\sqrt{v_5 v_6}}{2} \nonumber \\
&& b_{\pm}=\frac{w_1}{2}-\frac{\sqrt{w_6 w_5}}{2} \pm \frac{\sqrt{v_1 v_2}}{2} \mp \frac{\sqrt{v_5 v_6}}{2}  \nonumber \\
&& c_{+}=\frac{z_2}{z_1}\left( \frac{w_1}{2} \sqrt{\frac{w_6}{w_5}} +\frac{w_6}{2} -\frac{v_1}{2} \sqrt{\frac{v_6}{v_5}} -\frac{v_6}{2} \sqrt{\frac{v_1}{v_2}}  \right) \nonumber \\
&& c_{-}=\frac{z_1}{z_2}\left( \frac{w_1}{2} \sqrt{\frac{w_5}{w_6}} +\frac{w_5}{2} +\frac{v_2}{2} \sqrt{\frac{v_5}{v_6}} +\frac{v_5}{2} \sqrt{\frac{v_2}{v_1}}  \right) \nonumber \\
&& d_{+}=\frac{1}{z_1z_2}\left( \frac{w_1}{2} \sqrt{\frac{v_1 v_5}{v_2 v_6}} -\frac{v_1 w_5}{2 v_2} +\frac{v_5}{2} \sqrt{\frac{v_1}{v_2}} -\frac{v_1}{2} \sqrt{\frac{v_5}{v_6}}  \right) \nonumber \\
&& d_{-}={z_1 z_2}\left( \frac{w_1}{2} \sqrt{\frac{v_2 v_6}{v_1 v_5}} -\frac{v_6 w_5}{2 v_5} +\frac{v_2}{2} \sqrt{\frac{v_6}{v_5}} -\frac{v_6}{2} \sqrt{\frac{v_2}{v_1}}  \right) 
\end{eqnarray}
and from the above expressions we can indeed verify that the weights satisfy the 
free-fermion condition,
\begin{equation}
a_{+} a_{-} +b_{+} b_{-}-c_{+} c_{-} -d_{+} d_{-}=0
\end{equation}

At this point we recall that for toroidal boundary conditions 
the vertex 
weights $c_{\pm}$ and $d_{\pm}$ always occurs as product 
combinations $c_{+} c_{-}$ and $d_{+} d_{-}$ in the sums
of the partition function \cite{BAX}. This means that the partition 
function 
of the equivalent 
even eight-vertex model 
does not depend  on the free parameters $z_1$ and $z_2$ which can be
used to set $c_{-}=c_{+}$ and $d_{-}=d_{+}$. We next note that our mappings
between the mixed eight-vertex and the Ising model given by 
Eqs.(\ref{equia},\ref{equib})
fulfill the restriction (\ref{constra}) in the absence of a magnetic field.
As a consequence
of that we can map the Ising model with zero magnetic field onto the
even eight-vertex model with weights satisfying the free-fermion 
condition. Considering
the map defined by Eq.(\ref{equia}) we find that the respective 
weights of the free-fermion even
eight-vertex model are\footnote{Here we have set $\beta=1$}, 
\begin{eqnarray}
\label{free1}
&& a_{+}=2 \cosh(J_h) \cosh(J_v),~~a_{-}=2 \cosh(J_h) \sinh(J_v) \nonumber \\
&& b_{+}=2 \sinh(J_h) \cosh(J_v),~~b_{-}=2 \sinh(J_h) \sinh(J_v) \nonumber \\
&& c_{+}=c_{-}=\cosh(J_h) \sqrt{2 \sinh(2 J_v)} \nonumber \\
&& d_{+}=d_{-}=\sinh(J_h) \sqrt{2 \sinh(2 J_v)} 
\end{eqnarray}
while the equivalence given by Eq.(\ref{equib}) lead us to the following weights
\begin{eqnarray}
\label{free2}
&& a_{+}=2 \cosh(J_h) \cosh(J_v),~~a_{-}=2 \sinh(J_h) \sinh(J_v) \nonumber \\
&& b_{+}=2 \cosh(J_h) \sinh(J_v),~~b_{-}=2 \sinh(J_h) \cosh(J_v) \nonumber \\
&& c_{+}=c_{-}=d_{+}=d_{-}=\sqrt{\sinh(2J_h) \sinh(2J_v)} 
\end{eqnarray}

We emphasize that the above mappings are valid on a finite toroidal square 
lattice in which
the partition functions of Ising with zero magnetic field 
and the free-fermion eight-vertex models with weights (\ref{free1},\ref{free2})
are exactly the same. We point out that the equivalences here differ from the one
between the checkerboard Ising model and the even eight-vertex model with
the free-fermion condition \cite{BAX1}. In this case the mapping is 
valid only in the
thermodynamic limit and the partition function of the Ising model is
twice as that of the equivalent even free-fermion 
eight-vertex model. The extention
of such mapping to toroidal lattice requires to consider four types of
Ising model partition
functions combining periodic and antiperiodic 
boundary conditions \cite{DAV}. Therefore,
we believe that our mappings (\ref{free1},\ref{free2}) 
are new in the literature since the Ising model and
its equivalent free-fermion eight-vertex model are considered
on the same toroidal square lattice. 

\section{Integrable three-state spin model and the equivalent $\mathrm{R}$-matrix}

In this section we argue that our first spin-vertex correspondence
provides us in principle the means to determine the underlying $\mathrm{R}$-matrix of the
equivalent vertex model associated to a given integrable spin model. Recall that the $\mathrm{R}$-matrix 
is an essential object for the solution
of an integrable model by the quantum inverse scattering method.
The basic idea is to use the
mapping to built the explicit form of the Lax operator and afterwards 
we are left to solve the Yang-Baxter algebra for the 
respective $\mathrm{R}$-matrix. The Yang-Baxter algebra with a 
given Lax operator leads us to solve a set of linear relations  
for the $\mathrm{R}$-matrix elements avoiding us to deal with functional equations.
Here we shall discuss this alternative approach for an integrable three-state spin model and
for sake of illustration we choose the one with simplest weight structure. The example
is the $N=3$ Fateev-Zamolodchikov spin model \cite{FATZAMO} whose weights are given by,
\begin{equation}
W_h(x)=\left(
\begin{array}{ccc}
1 & b(x) & b(x) \\
b(x) & 1 & b(x) \\ 
b(x) & b(x) & 1 \\
\end{array}
\right),~~~
W_v(x)=\left(
\begin{array}{ccc}
1 & \bar{b}(x) & \bar{b}(x) \\
\bar{b}(x) & 1 & \bar{b}(x) \\ 
\bar{b}(x) & \bar{b}(x) & 1 \\
\end{array}
\right)
\end{equation}
where $b(x)=\frac{\sin\left(\frac{Pi}{6}-x\right)}{\sin\left(\frac{Pi}{6}+x\right)}$ and  
where $\bar{b}(x)=\frac{\sin\left(x\right)}{\cos\left(\frac{Pi}{6}+x\right)}$.  

We now use the spin-vertex correspondence (\ref{ansa}) to built the Lax
operator of the corresponding 27-vertex model. This operator can be represented by the following matrix,
\begin{equation}
\label{LAX27}
{\mathbb{L}}^{(27)}(x)=  \left[ 
\begin{array}{ccc|ccc|ccc}
w_1(x) & 0 & 0 & w_2(x) & 0 & 0 & w_2(x) & 0 & 0 \\
w_3(x) & 0 & 0 & w_4(x) & 0 & 0 & w_5(x) & 0 & 0 \\
w_3(x) & 0 & 0 & w_5(x) & 0 & 0 & w_4(x) & 0 & 0 \\ \hline
0 & w_4(x) & 0 & 0 & w_3(x) & 0 & 0 & w_5(x) & 0 \\
0 & w_2(x) & 0 & 0 & w_1(x) & 0 & 0 & w_2(x) & 0 \\
0 & w_5(x) & 0 & 0 & w_3(x) & 0 & 0 & w_4(x) & 0 \\ \hline
0 & 0 & w_4(x) & 0 & 0 & w_5(x) & 0 & 0 & w_3(x) \\
0 & 0 & w_5(x) & 0 & 0 & w_4(x) & 0 & 0 & w_3(x) \\
0 & 0 & w_2(x) & 0 & 0 & w_2(x) & 0 & 0 & w_1(x) \\
\end{array}
\right]
\end{equation}
where the vertex weights are given by,
\begin{eqnarray}
\label{wei27}
&& w_1(x) = 1,~~ 
w_2(x)=\frac{\sin\left(x\right)}{\cos\left(\frac{Pi}{6}+x\right)},~~ 
w_4(x)=\frac{\sin\left(\frac{Pi}{6}-x\right)}{\sin\left(\frac{Pi}{6}+x\right)} \nonumber \\
&& w_3(x)=w_5(x)=\frac{\sin\left(x\right)}{\cos\left(\frac{Pi}{6}+x\right)} 
\frac{\sin\left(\frac{Pi}{6}-x\right)}{\sin\left(\frac{Pi}{6}+x\right)}  
\end{eqnarray}

The next step in this approach is to find the $\mathrm{R}$-matrix which solves the Yang-Baxter algebra,
\begin{equation}
\label{RLL}
\mathrm{R}^{(27)}_{12}(x,y) {\mathbb{L}}^{(27)}_{13}(x){\mathbb{L}}^{(27)}_{23}(y)={\mathbb{L}}^{(27)}_{23}(y){\mathbb{L}}^{(27)}_{12}(x) \mathrm{R}^{(27)}_{12}(x,y)
\end{equation}

In order to determine the structure of the $\mathrm{R}$-matrix we choose two distinct point $x$ and $y$ and solve numerically 
the relation (\ref{RLL}) for a general $9 \times 9$ $\mathrm{R}$-matrix. By applying
this procedure for a number of distinct pair of points we find that many of the $\mathrm{R}$-matrix elements are zero. We conclude that the 
basic form of
the $\mathrm{R}$-matrix is similar to that of the Lax operator, namely
\begin{equation}
\label{RMA27}
\mathrm{R}^{(27)}(x,y)=  \left[ 
\begin{array}{ccc|ccc|ccc}
{\bf w}_1(x,y) & 0 & 0 & {\bf w}_2(x,y) & 0 & 0 & {\bf w}_2(x,y) & 0 & 0 \\
{\bf w}_3(x,y) & 0 & 0 & {\bf w}_4(x,y) & 0 & 0 & {\bf w}_5(x,y) & 0 & 0 \\
{\bf w}_3(x,y) & 0 & 0 & {\bf w}_5(x,y) & 0 & 0 & {\bf w}_4(x,y) & 0 & 0 \\ \hline
0 & {\bf w}_4(x,y) & 0 & 0 & {\bf w}_3(x,y) & 0 & 0 & {\bf w}_5(x,y) & 0 \\
0 & {\bf w}_2(x,y) & 0 & 0 & {\bf w}_1(x,y) & 0 & 0 & {\bf w}_2(x,y) & 0 \\
0 & {\bf w}_5(x,y) & 0 & 0 & {\bf w}_3(x,y) & 0 & 0 & {\bf w}_4(x,y) & 0 \\ \hline
0 & 0 & {\bf w}_4(x,y) & 0 & 0 & {\bf w}_5(x,y) & 0 & 0 & {\bf w}_3(x,y) \\
0 & 0 & {\bf w}_5(x,y) & 0 & 0 & {\bf w}_4(x,y) & 0 & 0 & {\bf w}_3(x,y) \\
0 & 0 & {\bf w}_2(x,y) & 0 & 0 & {\bf w}_2(x,y) & 0 & 0 & {\bf w}_1(x,y) \\
\end{array}
\right]
\end{equation}

By substituting the ansatz (\ref{RMA27}) and the expression for the Lax 
operator (\ref{LAX27},\ref{wei27}) in the Yang-Baxter algebra (\ref{RLL}) we find 
a number
of linear relations involving the $\mathrm{R}$-matrix elements. Many of these relations are 
linear dependent 
and we only need to 
solve four independent polynomial equations. As a result we obtain that 
the $\mathrm{R}$-matrix elements are,
\begin{eqnarray}
\label{weiRMA}
&& \frac{{\bf w}_2(x,y)}{{\bf w}_1(x,y)}=
\frac{\sin\left(x-y\right)}{\cos\left(\frac{Pi}{6}+x-y\right)}
\frac{\sin\left(\frac{Pi}{6}+y\right)}{\sin\left(\frac{Pi}{6}-y\right)} 
~~\frac{{\bf w}_3(x,y)}{{\bf w}_1(x,y)}=
\frac{\sin\left(x-y\right)}{\cos\left(\frac{Pi}{6}+x-y\right)}
\frac{\sin\left(\frac{Pi}{6}-x\right)}{\sin\left(\frac{Pi}{6}+x\right)} \nonumber \\
&& \frac{{\bf w}_4(x,y)}{{\bf w}_1(x,y)}=
\frac{\sin\left(\frac{Pi}{6}-x\right)}{\sin\left(\frac{Pi}{6}+x\right)} 
\frac{\sin\left(\frac{Pi}{6}+y\right)}{\sin\left(\frac{Pi}{6}-y\right)},~~ 
\frac{{\bf w}_5(x,y)}{{\bf w}_1(x,y)}=
\frac{\sin\left(x-y\right)}{\cos\left(\frac{Pi}{6}+x-y\right)}
\frac{\sin\left(\frac{Pi}{6}-x\right)}{\sin\left(\frac{Pi}{6}+x\right)} 
\frac{\sin\left(\frac{Pi}{6}+y\right)}{\sin\left(\frac{Pi}{6}-y\right)} \nonumber \\
\end{eqnarray}
and we observe that $\mathrm{R}$-matrix elements can not be written in terms of the difference of
the spectral parameters. Note also that for $y=0$ the $\mathrm{R}$-matrix reduces to the Lax operator (\ref{LAX27},\ref{wei27}).

We have verified that the $\mathrm{R}$-matrix $\mathrm{R}^{(27)}(x,y)$ satisfy the Yang-Baxter equation (\ref{YANBAX}) 
and if we choose the normalization ${\bf w}_1(x,y)=\frac{\sin(x-y) -\cos(\frac{\pi}{6})}{\sqrt{3}(\sin(x-y)-\sin(\frac{\pi}{6}))}$
we have the standard unitarity property, 
\begin{equation}
\mathrm{R}^{(27)}_{12}(x,y) \mathrm{R}^{(27)}_{21}(y,x)= \mathrm{I_9}
\end{equation}
reducing to the
$9 \times 9$ permutator at the point $y=x$.

We can explore the fact that the $\mathrm{R}$-matrix (\ref{RMA27},\ref{weiRMA}) is not of difference 
form to build an extention of the spin chain associated to the $N=3$ Fateev-Zamolodchikov 
spin model. This is done by defining the following transfer matrix,
\begin{equation}
\label{T27V}
T^{(27)}(x)= Tr_{{\cal{A}}}\left[ \mathrm{R}^{(27)}_{{\cal A} L}(x,x_0)  
\mathrm{R}^{(27)}_{{\cal A} L-1}(x,x_0) \dots  
\mathrm{R}^{(27)}_{{\cal A} 1 }(x,x_0) \right] 
\end{equation}
where the second spectral parameter plays the role of an additional independent
coupling of a generalized vertex model. 

The transfer matrix (\ref{T27V}) generates a family of local Hamiltonians because the regularity 
of the $\mathrm{R}$-matrix extends to all $x=x_0$. By expanding the logarithm of the transfer matrix (\ref{T27V}) around
the regular point $x=x_0$ we obtain, apart from multiplicative and additive constants, the following Hamiltonian,
\begin{eqnarray}
\label{H27}
H^{(27)} & =& -\sum_{j=1}^{L} \frac{2}{\sqrt{3}} \left (X_j +Z_j Z^{\dagger}_{j+1} +(X_j)^2+ (Z_j Z^{\dagger}_{j+1})^2 \right) \nonumber \\
&+& \frac{4 \sin(x_0)}{\sqrt{3}} \sum_{j=1}^{L} e^{-i(\frac{\pi}{6}+x_0)} \left( X_j Z_j Z^{\dagger}_{j+1} + ( X_j)^2 (Z_j Z^{\dagger}_{j+1})^2 \right) \nonumber \\
&+& \frac{4 \sin(x_0)}{\sqrt{3}} \sum_{j=1}^{L} e^{i(\frac{\pi}{6}+x_0)} \left( X_j (Z_j Z^{\dagger}_{j+1})^2 + (X_j)^2 Z_j Z^{\dagger}_{j+1} \right) 
\end{eqnarray}
where periodic boundary conditions are assumed and the operators $X$ and $Z$ denote the generators of the $Z_3$ symmetry,
\begin{equation}
X=\left(
\begin{array}{ccc}
0 & 0 & 1 \\
1 & 0 & 0 \\ 
0 & 1 & 0 \\
\end{array}
\right),~~~
Z=\left(
\begin{array}{ccc}
1 & 0 & 0 \\
0 & e^{\frac{2i\pi}{3}} & 0 \\ 
0 &  & e^{\frac{4i\pi}{3}} \\
\end{array}
\right)
\end{equation}

The first term of the Hamiltonian (\ref{H27}) is quantum spin chain associated 
to the $N=3$ Fateev-Zamolodchikov spin model \cite{ALLI,MCPER1,MCPER2} while the additional
interactions couple the generators of the $Z_3$ algebra. Recall here that this situation is similar to that
found in section 4 for the mixed eight-vertex model in which besides the Ising quantum chain we have the
extra Dzyaloshinky-Moriya interaction \footnote{Note that if we use the 
relation $\sigma^{y}=i \sigma^{x} \sigma^{z}$ the third term in Eq.(\ref{HAMMIXTOT1}) can be rewritten
as $\sigma_j^{x} \sigma_{j}^{z} \sigma_{j+1}^{z} $ which couples the generators of the $Z_2$ symmetry.} . It is plausible to believe that the above analysis can be extended
to include other integrable spin models such as the Chiral Potts model \cite{MCPER1,MCPER2}.
We hope to address to this problem as well as the analysis of the Yang-Baxter algebra of 
the $n$-state mixed vertex model with configurations defined by Eq.(\ref{mix}) in a future work.

\section{Conclusions}

In this paper we have presented evidences on the existence of two possible correspondences between $n$-state 
spin and vertex models 
on square lattice with periodic boundary conditions. 
The equivalences are in the sense that  
the partition functions of the spin and the vertex model coincide in a toroidal 
lattice with arbitrary
size. Essential to these mappings was to uncover the suitable vertex configurations of the equivalent vertex model which turns out to have
only $n^3$ non-null weights. From the point of view of algebraic
geometric such equivalences can be schematically represented by the following maps,
\begin{equation}
\renewcommand{\arraystretch}{1.5}
\begin{array}{ccc}
\mathrm{Spin}~\mathrm{Model} \subset P^{n^2-1} \times P^{n^2-1} & \overset{\varphi}{\longrightarrow} &  \mathrm{Vertex}~\mathrm{Model} \subset P^{n^3-1} \\
W_h(i_1,i_2), W_v(i_3,i_4) & \longmapsto & W_h(i_3,i_1) W_v(i_3,i_2) \delta_{i_1,i_4} \\
W_h(i_1,i_2), W_v(i_3,i_4) & \longmapsto & W_h(i_1,i_2) W_v(i_3,i_1) \delta_{i_1,i_4} 
\end{array}
\end{equation}

In particular, we have argued that the partition function of the Ising model 
in an external 
magnetic field 
can be reformulated
as the partition function of a mixed eight-vertex model. We have studied the 
Yang-Baxter relations 
for the mixed eight-vertex model with symmetrical weights and we 
find a solution 
lying on the
same elliptic curve associated to the even eight-vertex model uncovered 
by Baxter \cite{BAX}. The
elements of the $\mathrm{R}$-matrix associated to the symmetric mixed 
eight-vertex model can not however
be written in terms of the difference 
of spectral parameters parameterizing the Lax operators. In this sense 
the situation 
is distinct from 
that of the even
eight-vertex model in which the difference property is present 
in the underlying $\mathrm{R}$-matrix. In fact,
we have shown that the $\mathrm{R}$-matrix associated to the symmetric mixed 
eight-vertex model lie
on a quartic surface which is argued to be in the geometrical 
class of the $K3$ surfaces. The study of the underlying quantum spin chain
prompted us to investigate a mapping among the mixed eight-vertex model
and the even eight-vertex model with weights satisfying the free-fermion condition. 
As a consequence we have been able to propose
novel mappings among the Ising model in absence of a magnetic
field and the free-fermion even eight-vertex model which are valid 
for a toroidal lattice.

We have shown that mixed eight-vertex model with symmetric weights 
encodes the Ising model with zero magnetic field and thus such spin
model can in principle be tackled within 
the quantum inverse scattering framework. We think that this may be 
general situation of any spin model with commuting diagonal-to-diagonal
transfer matrices.
To this end we have to use 
the first spin-vertex correspondence to uncover the weights of the 
equivalent vertex model and after that one has to solve the vertex version 
of the Yang-Baxter 
algebra for the given Lax operator. This leads us to a set of linear
equations for the $\mathrm{R}$-matrix elements which are easier to solve than
typical functional relations involving both Lax operator 
and $\mathrm{R}$-matrix as unknown objects. As an example, we have applied this
method to determine the $\mathrm{R}$-matrix of the 27-vertex model whose 
partition function
is the same as that of the integrable three-state 
Fateev-Zamolodchikov spin model. The fact that the $\mathrm{R}$-matrix is not of 
difference form
can be used to generate an extention of the quantum spin chain of the three-state
Fateev-Zamolodchikov model in which the extra interactions couple the
$Z_3$ generators. The expectation is that such approach can be carried out to
other integrable spin models such as the Chiral Potts model \cite{MCPER1,MCPER2}.

We believe that our mapping may find other applications beyond
paving the way for finding common algebraic structures among spin and vertex models. 
We recall that the  vertex models have an intrinsic tensor 
structure amenable
for gauge transformation under which the partition 
function remains unchanged. This symmetry has been used to show that
that the partition function of any sixteen-vertex model can be expressed
in terms of set of irreducible polynomial algebraic invariants \cite{HIE,SHA,PERK3}.
In particular, these invariants have been used to locate the critical
line of the isotropic Ising model in a non-zero magnetic field \cite{WUWU}
exploring its relation to the sixteen-vertex mentioned in the introduction.
In this paper we have put forward a much simpler equivalent vertex model which 
covers the anisotropic Ising model with arbitrary horizontal and vertical ferromagnetic
couplings. Therefore, we expect that our mapping together
with the approach advocated in ref.\cite{WUWU} could be useful 
to determine critical frontier of the Ising model in a more generic situation.

\section*{Acknowledgments}

This work was supported in part by the Brazilian 
Research Council CNPq 305617/2021-4.

\addcontentsline{toc}{section}{Appendix A}
\section*{\bf Appendix A: The even eight-vertex model}
\setcounter{equation}{0}
\renewcommand{\theequation}{A.\arabic{equation}}

Here we summarize the Yang-Baxter analysis 
for the even eight-vertex 
weights with symmetrical
vertex weights. Following the notation 
used in Baxter 
monograph \cite{BAX} we set,
\begin{equation}
w_1=w_2=a,~~w_3=w_4=b,~~w_5=w_6=c,~~w_7=w_8=d
\end{equation}

The corresponding matrices representation 
for the Lax operator 
and the $\mathrm{R}$-matrix are,
\begin{equation}
\label{LAXRMA}
\mathbb{L}^{(\mathrm{even})}(\omega)=\left[
\begin{array}{cc|cc}
a & 0 & 0 & d \\
0 & b & c & 0 \\ \hline
0 & c & b & 0 \\
0 & 0 & 0 & a \\
\end{array}
\right],~~~
\mathrm{R}^{(\mathrm{even})}(\omega^{'},\omega^{''})=\left[
\begin{array}{cc|cc}
{\bf a} & 0 & 0 & {\bf d} \\
0 & {\bf b} & {\bf c} & 0 \\ \hline
0 & {\bf c} & {\bf b} & 0 \\
{\bf d} & 0 & 0 & {\bf a} \\
\end{array}
\right],
\end{equation}

Baxter has shown that these pair of matrices satisfy the
Yang-Baxter algebra (\ref{YBAV}) 
provided that the vertex weights  lie
on the intersection of the following quadrics, namely
\begin{equation}
\label{INVB1}
\frac{c^{'} d^{'}}{a^{'} b^{'}}=
\frac{c^{''} d^{''}}{a^{''} b^{''}}= \Delta_1=\frac{1-\Gamma}{1+\Gamma} 
\end{equation}
and
\begin{equation}
\label{INVB2}
\frac{(a^{'})^2+(b^{'})^2-(c^{'})^2-(d^{'})^2}{2 a^{'} b^{'}}=
\frac{(a^{''})^2+(b^{''})^2-(c^{''})^2-(d^{''})^2}{2 a^{''} b^{''}}= \Delta_2=\Delta(1+\Gamma)
\end{equation}
where $\Gamma$ and $\Delta$  denote the constant parameter 
originally used by Baxter \cite{BAX}.

Comparing the algebraic invariants of the 
mixed eight-vertex (\ref{INV1},\ref{INV2}) 
with those associated to the even eight-vertex 
model (\ref{INVB1},\ref{INVB2}) we observe
the immediate correspondence, 
\begin{equation}
\label{MAP}
w_1=a,~~v_5=b,~~,w_5=c,~~v_1=d
\end{equation}
and by using Baxter's parameterization 
of the weights $a$, $b$, $c$, $d$ 
we obtain the uniformization given in
Eq.(\ref{UNIF}).

By way of contrast the expressions of the 
elements of even 
eight-vertex $\mathrm{R}$-matrix  
are quite different from that of the mixed 
eight-vertex model given 
in Eqs.(\ref{v0w1v0v1},\ref{v0v5}). In fact, choosing
the entry ${\bf c}$ as an overall
normalization one finds, in the notation of this paper, 
the following results,
\begin{eqnarray}
\label{RMABAX}
&& \frac{{\bf a}}{{\bf c}}=\frac{(a^{''})^2}{(c^{''})^2}
\frac{\left(c^{'} c^{''}-d^{'} d^{''}\right) \Big((c^{'})^2 (b^{''})^2-(a^{'})^2 (c^{''})^2\Big)}{\left(b^{'} b^{''}-a^{'} a^{''}\right) \Big((c^{'})^2 (a^{''})^2-(a^{'})^2 (d^{''})^2\Big)} \nonumber \\
&& \frac{{\bf b}}{{\bf c}}=
\frac{a^{''} b^{''}}{c^{''} d^{''}}
\frac{(c^{'} d^{''}-d^{'} c^{''})}{(b^{'} b^{''}-a^{'} a^{''})} \nonumber \\
&& \frac{{\bf d}}{{\bf c}}=
\frac{d^{''} a^{''}}{b^{''} c^{''}}
\frac{\left(b^{'} a^{''}-a^{'} b^{''}\right) \Big((c^{'})^2 (b^{''})^2-(a^{'})^2 (c^{''})^2\Big)}{\left(b^{'} b^{''}-a^{'} a^{''}\right) \Big((c^{'})^2 (a^{''})^2-(a^{'})^2 (d^{''})^2\Big)}
\end{eqnarray}

By comparing the $\mathrm{R}$-matrix elements (\ref{RMABAX}) 
with those 
associated to 
the mixed eight-vertex model (\ref{v0w1v0v1},\ref{v0v5}) 
we conclude that the Lax operator 
mapping (\ref{MAP})
does not extend to the $\mathrm{R}$-matrix.  This difference can be further 
emphasized by computing the 
algebraic invariants (\ref{INVB1},\ref{INVB2}) replacing the Lax weights 
by the $\mathrm{R}$i-matrix elements, namely 
\begin{equation}
\label{INV1Bv0}
\frac{{\bf c} {\bf d}}{{\bf a} {\bf b}}=G_1(a^{'},\dots,d^{'},a^{''},\dots,d^{''}),~~
\frac{{\bf{a}}^2+{\bf b}^2-{\bf c}^2-{\bf d}^2}{2 {\bf a} {\bf b}}
=G_2(a^{'},\dots,d^{'},a^{''},\dots,d^{''})
\end{equation}
and
by using systematically the algebraic invariants (\ref{INVB1},\ref{INVB2}) for 
the Lax vertex weights we obtain
\begin{equation}
\label{INV2Bv0}
G_1(a^{'},\dots,d^{'},a^{''},\dots,d^{''})=\Delta_1,~~
G_2(a^{'},\dots,d^{'},a^{''},\dots,d^{''})=\Delta_2
\end{equation}

As a consequence of Eqs(\ref{INV1Bv0},\ref{INV2Bv0}) we see that the 
$\mathrm{R}$-matrix lie on the 
same elliptic curve of the Lax operators. This means that 
the $\mathrm{R}$-matrix elements provide the group or addition 
law on the elliptic curve defined by the Lax vertex weights. This is the 
geometrical reason why
the $\mathrm{R}$-matrix of the symmetric eight-vertex model may be
expressed in terms of the difference of spectral parameters. 

\addcontentsline{toc}{section}{Appendix B}
\section*{\bf Appendix B: Elimination procedure}
\setcounter{equation}{0}
\renewcommand{\theequation}{B.\arabic{equation}}

In order to obtain the expression for the surface we have to eliminate 
the weights $w_1^{''}$, $w_5^{''}$, $v_1^{''}$ and 
$v_5^{''}$ from the relations 
(\ref{INVdois},\ref{FUNdois}). Recall here that such weights are constrained
by the invariants,
\begin{equation}
\frac{w^{''}_5 v^{''}_1}{w^{''}_1 v^{''}_5}= \Delta_1,~~
\frac{(w^{''}_1)^2+(v^{''}_5)^2-(w^{''}_5)^2-(v^{''}_1)^2}{2 w^{''}_1 v^{''}_5}= \Delta_2
\end{equation}

We can use the invariant $\Delta_1$ to eliminate for instance 
the weight $v^{''}_1=\Delta_1 \frac{w^{''}_1 v^{''}_5}{w^{''}_5}$ and we substitute this weight 
in  Eqs.(\ref{INVdois},\ref{FUNdois}).  With
the help of the invariant $\Delta_2$ and after some simplifications we obtain,
\begin{equation}
\label{ELI1}
\frac{{\bf w}_5 {\bf v}_1}{{\bf w}_1 {\bf v}_5}= \Delta_1 \left(\frac{w^{''}_1}{w^{''}_5}\right)^2,~~
\frac{({\bf w}_1)^2+({\bf v}_5)^2-(\bf{w_5})^2-({\bf v_1})^2}{2 {\bf w_1} {\bf v_5}}=\Delta_1 \frac{w^{''}_1 v_1^{''}}{(w^{''}_5)^2}
+\frac{\Delta_2 w^{''}_1-v^{''}_5}{w^{''}_5}
\end{equation}
while the other weights 
$w_1^{''}$, $w_5^{''}$ and $v_5^{''}$  lie on the 
following quartic curve,
\begin{equation}
\label{ELI2}
(w^{''}_5)^2\left((v^{''}_5)^2+(w^{''}_1)^2-(w^{''}_5)^2-2\Delta_2 w^{''}_1 v^{''}_5\right)-\left(\Delta_1 v^{''}_5 w^{''}_1\right)^2=0
\end{equation}

The polynomial (\ref{ELI2}) is homogeneous and therefore we can carry on the elimination procedure 
defining affine coordinates such as
$x=\frac{w^{''}_1}{w^{''}_5}$ and 
$y=\frac{v^{''}_5}{w^{''}_5}$. In terms of these affine variables Eqs.(\ref{ELI1},\ref{ELI2}) becomes,
\begin{eqnarray}
\label{ELI3}
&& ({\bf v}_5)^2 - ({\bf v}_1)^2 + ({\bf w}_1)^2 - ({\bf w}_5)^2 - 2 \Delta_2 {\bf v}_5 {\bf w}_1 x + 
     2 {\bf v}_5 {\bf w}_1 y - 2 (\Delta_1)^2 {\bf v}_5 {\bf w}_1 x^2 y=0 \\
\label{ELI4}
&& {\bf v}_1 {\bf w}_5 - \Delta_1 {\bf v}_5 {\bf w}_1 x^2=0 \\
\label{ELI5}
&& x^2 - 2 \Delta_2 x y + y^2 - (\Delta_1)^2 x^2 y^2-1=0
\end{eqnarray}

From Eq.(\ref{ELI3}) we can eliminate the affine variable y,
\begin{equation}
y = \frac{({\bf v}_5)^2 - ({\bf v}_1)^2 + ({\bf w}_1)^2 - ({\bf w}_5)^2 - 2 \Delta_2 {\bf v}_5 {\bf w}_1 x}{
     2 {\bf v}_5 {\bf w}_1 \left(-1 + (\Delta_1)^2 x^2\right)}
\end{equation}
and after substituting this variable in Eq.(\ref{ELI5}) we find,
\begin{eqnarray}
\label{final}
&& -({\bf v}_1)^4 - ({\bf v}_5)^4 - \Big(({\bf w}_1)^2 - ({\bf w}_5)^2\Big)^2 + 
     2 {\bf v}_1^2 \Big(({\bf v}_5)^2 + ({\bf w}_1)^2 - ({\bf w}_5)^2\Big)  \nonumber \\
&&     +2 ({\bf v}_5)^2 \Bigg(({\bf w}_5)^2 + ({\bf w}_1)^2 \Big(1 + 2 x^2 \big(-1 + (\Delta_2)^2 + 
           (\Delta_1)^2 (-1 + x^2)\big)\Big)\Bigg)=0
\end{eqnarray}

Finally, we note that the expression (\ref{final}) depends on the last affine variable as $x^2$. This power can be
easily eliminate with the help of Eq.(\ref{ELI4}) leading us to the quartic surface (\ref{SURF}).

\end{document}